\documentclass[graybox, secnum]{svmult}

\usepackage{mathptmx}       
\usepackage{helvet}         
\usepackage{courier}        
\usepackage{type1cm}        
\usepackage{makeidx}         
\usepackage{graphicx}        

\usepackage{subfig}
\graphicspath{{./figures/}}
\usepackage{multicol}        
\usepackage[bottom]{footmisc}
\usepackage{hyperref}        
\usepackage{soul}            
\hypersetup{colorlinks=true,urlcolor=blue}
\usepackage{siunitx}        
\usepackage[square,numbers]{natbib}
\newcommand*\chem[1]{\ensuremath{\mathrm{#1}}}
\makeindex             

\begin{document}
\tableofcontents{}
\title*{Filters for X-ray detectors on Space missions}
\author{Marco Barbera\thanks{corresponding author} and Ugo Lo Cicero and Luisa Sciortino}
\institute{Marco Barbera \at Università degli Studi di Palermo, Dipartimento di Fisica e Chimica "E. Segrè", Viale delle Scienze, Ed. 17, 90128 Palermo, Italy, \email{marco.barbera@unipa.it}
\and Ugo Lo Cicero \at Istituto Nazionale di Astrofisica, Osservatorio Astronomico di Palermo "G. S. Vaiana", Piazza del Parlamento 1, 90134 Palermo, Italy, \email{ugo.locicero@inaf.it}
\and Luisa Sciortino \at Università degli Studi di Palermo, Dipartimento di Fisica e Chimica "E. Segrè", Viale delle Scienze, Ed. 17, 90128 Palermo, Italy, \email{luisa.sciortino@unipa.it}}
%
%
\maketitle
\abstract{Thin filters and gas tight windows are used in Space to protect sensitive X-ray detectors from out-of-band electromagnetic radiation, low-energy particles, and molecular contamination.
Though very thin and made of light materials, filters are not fully transparent to X-rays. For this reason, they ultimately define the detector quantum efficiency at low energies.
In this chapter, we initially provide a brief overview of filter materials and specific designs adopted on space experiments with main focus on detectors operating at the focal plane of grazing incidence X-ray telescopes. We then provide a series of inputs driving the design and development of filters for high-energy astrophysics space missions. We begin with the identification of the main functional goals and requirements driving the preliminary design, and identify modeling tools and experimental characterization techniques needed to prove the technology and consolidate the design. Finally, we describe the calibration activities required to derive the filter response with high accuracy. We conclude with some hints on materials and technologies presently under investigation for future X-ray missions.}


\section{Introduction}
\label{Intro}

In order to fully exploit the capabilities of X-ray detectors operating in Space missions, thin filters, transparent to the energy range of interest, need to be used to protect the sensitive detector area from out-of-band electromagnetic radiation, low energy particles, and molecular contamination or to reduce count rates for very bright sources. 

Submicron thickness polymer foils coated with a low Z metal have been traditionally used to respond to such requirements. The metal blocks the UV/VIS/IR light, mainly by reflection, and shapes the bandpass in the soft X-ray region of the spectrum, while the polymer serves as structural support for the ultra thin metal foil and partially attenuates the UV region. 

The main X-ray detectors used in high-energy astrophysics space missions are: gas proportional counters, microchannel plates, semiconductor detectors and, more recently, microcalorimeters (see different chapters in this volume). Each detector technology has different requirements on the filters, however, in any case, X-ray transparency shall be maximized while still providing adequate out-of-band rejection and mechanical resistance to the vibro-acoustic launch stresses.

Gas proportional counters (PC), used as soft X-ray detectors since the early stage of space exploration, need thin windows to confine the gas in the ionization chamber. Such windows, usually supported by a metal mesh to withstand a few hundreds millibars of gas pressure, are usually not fully leak-tight and a gas flow is needed to replenish the gas mixture to maintain a constant gain. The thin window is typically coated with aluminium or carbon to be electrically conductive, and also to reduce the UV-induced photoelectrons within the time window of the X-ray photon detection pulse. 

Microchannel plate (MCP) X-ray detectors have high sensitivity to UV photons as well as to low-energy charged particles. The deposition of CsI or KBr photocathode on the front face of an MCP considerably enhances the detector quantum efficiency (QE) in the soft X-ray, but also pushes its UV sensitivity out beyond 200~nm. A thin filter with conductive coating needs to be mounted close to the sensitive detector area to block the diffuse UV background, mainly due to solar radiation scattered by the interplanetary medium, and bright UV sources (mainly hot stars) \cite{Barbera1994}. A small voltage is usually applied to the filter to prevent low-energy charged particles to hit the front MCP and generate background signals.

An experimental investigation conducted on polycarbonate/Al filters, initially designed for the Chandra HRC \cite{Barbera1997}, has shown that the occurrence of Fabry-Perot interference effects, due to the multilayer design of the filter, and the oxidation of aluminium cause a two orders of magnitude increase in UV transmission at $\sim$200~nm with respect to the prediction based on a simple slab model of transmission. The results of this work have led to a redesign of the HRC UV/Ion shields replacing polycarbonate with polyimide (PI) and depositing the Al layer on one single side \cite{Murray1997,Meehan1997}. The same results have also stimulated the redesign of the XMM-Newton EPIC CCD (charge coupled device) filters \cite{Villa1998}, the JET-X filters \cite{Castelli1997}, and the filters for the Chandra ACIS CCD \cite{Powell1997,Garmire2003}.

Semiconductor detectors, such as CCDs, silicon drift detectors (SDD), and DEPFET active pixel arrays are sensitive to UV (ultraviolet)/VIS (visible) photons with energy larger than the Si band-gap ($\approx$1.1~eV). The detection of UV/VIS photons degrades the detector spectral resolution and changes the energy scale by about 3.7~eV, on average, for each optically generated electron-hole pair during the integration time window. For this reason, ideally, no optically generated electron-hole pair should be detected during a read-out integration time. An optical blocking filter (OBF) is thus needed to remove UV/VIS background light and to allow the observation of X-ray sources with bright UV/VIS counterparts.

Most semiconductor detectors flown to date have used free-standing filters (Chandra ACIS \cite{Garmire2003}, XMM-Newton EPIC-MOS \cite{Turner2001}, XMM-Newton EPIC-PN \cite{Struder2001}, Swift XRT \cite{Burrows2004}, Suzaku XIS \cite{Koyama2007}). Direct deposition of an optical blocking filter on the photon entrance window of the
sensor chip has been implemented on a few experiments such as the CCDs of the reflection grating spectrometer (RGS) experiment on board XMM-Newton \cite{denherder2001}, the soft x-ray imager (SXI) on board the Hitomi x-ray observatory \cite{Tanaka2018}, the REgolith
X-ray Imaging Spectrometer (REXIS) on board the NASA’s OSIRIS-REx asteroid sample return mission \cite{Bautz2016}\cite{Thayer2021}, and on five of the seven CCDs of the eROSITA mission launched in 2019 \cite{Meidinger2021}. Some of the above experiments have reported light leakages due to pinholes on the on-chip Al layer OBFs.  In the case of the Athena Wide Field Imager (WFI) \cite{Meidinger2014}, in order to not significantly deteriorate the low energy response, a mixed configuration is baselined with a filter deposited on-chip to get rid of UV/VIS background light and an additional OBF on the filter wheel, that will be used for the observation of X-ray sources with bright UV/VIS counterparts \cite{Barbera2018b}. 

Microcalorimeters, providing unprecedented energy dispersive spectral resolution, are considered among the most promising X-ray detectors for future high-energy astrophysics missions. The quite well developed array technology \cite{Bandler2019} makes them also valuable to perform spatially resolved spectral analysis of extended sources. Microcalorimeters have been already tested successfully on rocket experiments \cite{McCammon2002}, and on board the Hitomi Japanese space mission \cite{Kelley2016}, though only during a short phase of commissioning before the satellite was lost. Microcalorimeters will be operated on board the Japanese mission XRISM \cite{Tashiro2020}, on board Athena \cite{Barcons2015}, the next large mission of the European Space Agency Cosmic Vision science program to be launched in the early '30s, and finally, they are investigated for Lynx, a large NASA mission concept currently under study in the decadal survey 2020  \cite{Gaskin2019} and on the Hot Universe Baryon Surveyor (HUBS) mission in the Chinese space program \cite{Cui2020}. 

Microcalorimeters operate at very low temperature (T $\sim$ 100~mK) inside a sophisticated cryostat. In order to allow the X-rays to reach the detector, a clear path has to be opened in the cryostat thermal and structural shields surrounding the cold stage. A set of thin filters, highly transparent to X-rays, need to be mounted on the shield openings to reduce the IR (infrared) radiation heat-load from warm surfaces and to prevent degradation of the energy resolution by photon shot noise due to out-of-band low energy radiation. The filters mounted on the cryostat shields also protect the cold detector from molecular and particle contamination, and in some cases need to attenuate the radiofrequency (RF) electromagnetic interference from telemetry and spacecraft operation. The filters also significantly reduce the out-of-band radiation from target sources and background in the telescope field of view (FOV), however, particularly bright sources in the UV/VIS (e.g. massive O stars, planets, quasars, etc.) may need the use of additional optical blocking filters, typically mounted on a filter wheel located outside the cryostat vacuum enclosure. 

Some detectors, with insufficient counting rate dynamic range or in order to avoid pile-up of X-ray photons in a pixel, may need filters to attenuate particularly x-ray bright sources. Neutral Density micro etched metal foils with few \% transmission are used to uniformly attenuate the full spectrum, while few tens of $\mu$m thick Be filters or Al coated Kapton foils are used to suppress the low energy events allowing the detection of high energy photons (e.g. Fe-K line complex). An example is the SXS instrument on Hitomi \cite{DeVries2017}, or Resolve on XRISM where the spectral resolution of the micro-calorimeter detectors would significantly degrade if too many photons fell on the detector pixels.

Besides filters specifically designed to allow proper operation of X-ray detectors in space, prefilters, directly mounted at the entrance pupil of the telescopes, are sometimes used in space experiments. In particular, prefilters are needed in experiments designed to observe the solar chromosphere and the corona in order to reduce the direct Sun visible light and the heat load onto the instrument. The most used material for this purpose is aluminium. When exposed to full Sun in space, prefilters will become hot and proper thermal conduction over the filter to the support frame need to be guaranteed to cool the filter. For this purpose, and to strengthen the filter against launch loads, the aluminium can be mounted on thick metal meshes (e.g. Hinode XRT experiment \cite{Golub2007}, Solar Orbiter EUI experiment \cite{Rochus2020}). 
Prefilters mounted directly onto the telescope can be also needed as X-ray transparent thermal shields, to keep an X-ray telescope at proper uniform operating temperature reducing the power needed to balance the radiative loss towards the cold space (e.g. Symbol-X \cite{Collura2009}) or to control the mirror temperature in low Earth orbits where the satellite is subject to significant temperature gradients while alternating on hourly timescales between solar and Earth irradiation (e.g. Hitomi \cite{Takahashi2016}). 

Finally, bandpass filters can be used to recover some spectral information on the detected sources when using X-ray detectors with poor energy resolution (e.g. microchannel plates) or detectors not operating in single-photon counting mode. A combination of different materials (e.g. C, Be, B, Al, Ti, Sb, Sn, …) and thicknesses are used to properly define the bandpass, around the electron binding energies of the filter material, and/or high-energy pass filters (e.g. Hinode XRT experiment \cite{Golub2007}, EUE experiment \cite{Vedder1989}, EXOSAT LET instrument \cite{Paerels1990}). 

Prefilters and bandpass filters will not be discussed further, being beyond the main focus of this chapter. Section \ref{Overview} describes the adopted filter configurations for the X-ray detectors on board the main high energy astrophysics space satellites. Section \ref{Functional_goal} lists the functional goals of filters, and section \ref{Requirements} provides the main requirements driving their design. Section \ref{Mat_Tech} describes the most used filter materials and technologies in space, section \ref{Performance_modeling} describes performance modeling supporting the design. Section \ref{Characterization} describes some of the characterization techniques used to verify filter performances during the development and technology assessment phases, while the specific activities needed to calibrate flight models are discussed in section \ref{Calibration}. Finally, in section \ref{Future}, we mention some of the materials and technologies presently under investigation for future missions.

\section{Overview of filters on Space X-ray observatories}
\label{Overview}
A comprehensive survey of all the filters employed to operate X-ray detectors on board of space missions will be too extensive, therefore, we decided to describe in this chapter only the main experiments operating at the focal plane of soft X-ray focusing optics.  

The NASA \textbf{Einstein} satellite \cite{Giacconi1979}, launched into low Earth orbit (perigee 465~km, apogee 476~km, inclination \SI{23.5}{\degree}) in 1978, was the first satellite to carry on a real focusing X-ray telescope with different detectors selectable at the telescope focal plane. The main detectors were the Imaging Proportional Counter (IPC) with multiwire anodes\cite{Gorenstein1981} and the High-Resolution Imager (HRI) based on microchannel plates \cite{Henry1977}. 
The two IPC modules used different entrance windows: polypropylene \SI{2}{\micro\metre}/Lexan \SI{0.4}{\micro\metre} (module A), and Mylar \SI{3}{\micro\metre} (module B), respectively, both coated with \SI{0.2}{\micro\metre} of carbon to make the windows electrically conductive and also to reduce the UV induced photo-electrons within the time window of the X-ray photon detection pulse. 
The three HRI modules were identical except for the material used for the UV-ion shield placed in front of the MCP’s, in particular: \SI{1.13}{\micro\metre} parylene N + \SI{0.126}{\micro\metre} Al (Detector \#1), \SI{1.48}{\micro\metre} parylene N + \SI{0.051}{\micro\metre} Al (Detector \#2), \SI{0.72}{\micro\metre} parylene N + \SI{0.054}{\micro\metre} Al (Detector \#3).

\textbf{EXOSAT}, the first European Space Agency mission dedicated to high-energy astrophysics \cite{Taylor1981}, was launched into a highly eccentric near Earth orbit (Perigee 347~km, apogee 192.000~km, inclination \SI{72.5}{\degree}) in May 1983 and operated until May 1986. One of the main instruments on board EXOSAT was the low-energy imaging telescope \cite{deKorte1981}, consisting of two 1~m focal length Wolter I type grazing incidence optics, each one equipped with two focal plane detectors, a position sensitive proportional counter (PSD), and a channel multiplier array (CMA) mounted on an exchange mechanism. The PSD main entrance window consisted of a polypropylene membrane \SI{80}{\nano\metre} thick coated with C and supported by a 65\% transmission mesh, in addition, a second window was placed close to the main entrance window consisting of \SI{0.5}{\micro\metre} thick polypropylene coated with \SI{0.4}{\micro\metre} of Lexan.
Five filters could be selected in front of both detectors, namely: \SI{2.5}{\micro\metre} Teflon, \SI{0.5}{\micro\metre} polypropylene + \SI{1}{\micro\metre} boron, \SI{0.3}{\micro\metre} Lexan, \SI{0.4}{\micro\metre} Lexan, and \SI{0.1}{\micro\metre} parylene + \SI{0.1}{\micro\metre} aluminium. The last three are mainly designed to protect the microchannel plate detector from UV sky background at 58.4~nm (He I) and 121.6~nm (H Ly-$\alpha$). The CMAs are sensitive to ultraviolet photons and this caused contamination for observation of bright O and B stars, pointed or serendipitous in the field. The observations obtained with the boron/polypropylene filter were free of UV contamination and the aluminium/parylene is relatively immune to it except for fields with the very brightest and earliest stars. Observation taken with the Lexan filters suffered a more significant UV contamination (see HEASARC archive reports).

The German satellite \textbf{ROSAT} \cite{Truemper1982}, launched into low Earth orbit (perigee 580~km, apogee~580 km, inclination \SI{53}{\degree}) in 1990, strongly relied on the Einstein heritage and was equipped with two interchangeable detectors at the telescope focal plane of the same technology as the IPC and HRI, but with increased sensitivity and FOV. The Position Sensitive Proportional Counter (PSPC) \cite{Briel1995} used a window of polypropylene \SI{1}{\micro\metre} /Lexan \SI{0.4}{\micro\metre} coated with \SI{50}{\micro\gram\per\centi\metre\squared} of carbon, while the High-Resolution Imager (HRI) \cite{Pfeffermann1986}\cite{Zombeck1990} was equipped with two metalized plastic membranes: \SI{270}{\nano\metre} of polypropylene + \SI{30}{\nano\metre} of aluminium on one side, which served as electrostatic shield close to the front MCP, and \SI{640}{\nano\metre} Lexan coated on both sides with aluminium, to a total thickness of 60~nm, to block the UV background and bright source radiation. The choice of polypropylene for the electrostatic shield was driven by the need to get good transparency to the UV calibration light. 

Early observations with the ROSAT HRI exhibited an unexpected excess in the count rates of a number of A-type stars with respect to predictions based on the ROSAT Position Sensitive Proportional Counter (PSPC) observations and the measured HRI soft X-ray sensitivity. The detected UV contamination was fully understood during the Chandra development program through experimental tests conducted on filter samples investigated for the HRC \cite{Barbera1997} \cite{Zombeck1997}, and allowed to determine the effective sensitivity of the ROSAT HRI to ultraviolet radiation by the analysis of observations conducted on stars of different spectral types\cite{Barbera2000}.

The Japanese spacecraft Astro-D \cite{Tanaka1994}, launched into low Earth orbit (perigee 524~km, apogee 615~km, inclination 31.1°) in early 1993, and renamed \textbf{Asuka} (flying bird, then shortened to ASCA for Advanced Satellite for Cosmology and Astrophysics), was equipped with four large collecting area and moderate angular resolution thin foil optics. Two were focused on CCD arrays (Solid-state Imaging Spectrometer, SIS) and two on imaging gas scintillation proportional counters (Gas Imaging Spectrometer, GIS) \cite{Ohashi1996}. The GIS gas cell entrance window consisted of a \SI{10}{\micro\metre} thick beryllium foil, 66~mm outer diameter and 52~mm clear aperture diameter, bonded with high-temperature epoxy to a cupronickel flange. In order to withstand more than 1~atm of differential pressure, the window was supported by a copper-plated thin molybdenum grid, and a stainless steel fine mesh coated with tin. An optical blocking filter was placed in front of the SIS CCD, consisting of 100~nm thick unsupported film of Lexan coated on both sides with aluminium, each layer 40~nm thick.

\textbf{BeppoSAX}, an Italian program with the participation of the Netherlands, was launched into a low Earth orbit (perigee 575~km, apogee 594~km, inclination \SI{4}{\degree}) in 1996 with on board several moderate size instruments to perform observations over a broad energy range 0.1–200~keV. The soft X-ray band (E~$<$~10~keV) was covered by a set of four mirrors, double cone approximations to Wolter I geometry, with imaging gas scintillation proportional counter (GSPC) detectors located at the focal planes \cite{Parmar1997}. The low energy response of the GSPC’s was modulated by three optical elements on the focused beam path. Two of them protected the detector against space plasma, a fine-pitch Au-coated tungsten grid located at the exit of the mirror unit, and a protection window located just beneath the shutter of the aluminium detector enclosure. The protection window consisted of a multilayer AlN 20~nm/polyimide 40~nm/AlN 10~nm/polyimide 160~nm/carbon 2~nm mechanically supported by a \SI{250}{\micro\metre} pitch hexagonal polyimide grid. The third element was the gas cell entrance window which consisted of three layers of polyimide separated by Al/AlN multilayers, for a total of \SI{1.25}{\micro\metre} of polyimide, 44~nm of Al and 35~nm of AlN. This design was chosen to minimize the leak rate and the sensitivity to atomic oxygen corrosion.

The NASA Advanced X-ray Astrophysics Facility (AXAF, renamed \textbf{Chandra}) was placed in near-Earth orbit (perigee 14300~km, apogee 135000~km, inclination \SI{76.7}{\degree}) in July 1999 and is still operational. It was designed to have four times the effective area of Einstein at low energies and a considerable collecting area between 6~keV and 7~keV. The key elements of the mission were the high angular resolution (0.5~arcsec FWHM) Wolter type I geometry grazing incidence X-ray optics, and four selectable detectors at the focal plane: the Advanced CCD Imaging Spectrometer (ACIS) \cite{Garmire2003} consisting of four CCD chips in a 2x2 imaging array (16’ x 16’ FOV) and 6  CCD chips in a linear array for high-energy dispersive spectroscopy (0.4-10~keV), and the High-Resolution Camera (HRC) \cite{Zombeck1995} also consisting of two detectors based on microchannel plates, the imaging HRC-I (30’ x 30’ FOV) and the linear HRC-S for soft X-ray dispersive spectroscopy (0.07-0.15~keV). An Optical Blocking Filter (OBF) was placed about 2~cm above the CCDs to reject UV/VIS light from hot stars and from scattered light in the spacecraft. The filters consisted of a free-standing polyimide film 200~nm thick substrate coated with 160~nm of Al for ACIS-I and 130~nm of Al for the ACIS-S, where less visible light was expected thanks to the grating dispersion. The HRC was equipped with a UV/Ion shield, placed a few millimeters above the front MCP, consisting of a free-standing polyimide film $\sim$550~nm thick coated with 76~nm of Al for the HRC-I and, for the HRC-S, a slightly thinner polyimide film ranging between 210~nm and 275~nm with Al coating between 30~nm and 75~nm thick depending on the portion of the spectroscopic detector (see \cite{Murray2000} for a more detailed description of the HRC-S filter layout). 
During the Chandra in-flight calibrations, the star Vega (A0V, V = 0.03) was observed with both the HRC-I and the HRC-S to verify the efficiency of the UV/Ion shields. The predicted and measured counting rates were in good agreement. UV contamination was reduced by approximately a factor of one hundred compared to the ROSAT HRI detector \cite{Kenter2000}. 
On the ACIS CCD detector \cite{Garmire2003} of Chandra, a loss of sensitivity at low energy was detected since the first operating years, soon associated with a molecular contamination layer building up at the surface of the OBFs facing the spacecraft interior \cite{Plucinsky2016}. The main contaminants were identified by on board X-ray analysis as carbon, oxygen, and fluorine.  Even though the exact molecular source is still under debate, the most likely sources of contaminant are carbonaceous compounds, fluorocarbon compounds, and water \cite{Marshall2004}. An estimate for the accumulated molecular contamination is of 250–500 \si{\micro\gram\per\centi\metre\squared} of unidentified molecular contamination over 18 years of operation, which is one hundred times the pre-flight estimates and ten times the amount of carbon present in the OBFs polyimide \cite{ODell2017}.

The European X-ray Multi-Mirror Mission (XMM, renamed \textbf{XMM-Newton}) was launched in near-Earth orbit (perigee 5600~km, apogee 113000~km, inclination \SI{67.1}{\degree}) in late 1999 and is still operational. Like Chandra, it was designed to perform both imaging and spectroscopy by use of arrays of CCD chips and reflection gratings. The observatory consisted of three co-aligned high effective areas multi-shell Wolter type I X-ray telescopes, each one consisting of 58 nested Au-coated nickel electro-formed shells. Each X-ray telescope has a CCD at the focal plane, two of them are metal-oxide semiconductor (MOS) CCDs (operating in combination with the reflection gratings) \cite{Turner2001} and the third one is a pn-CCD used solely for spectral imaging \cite{Struder2001}. Three different filters can be selected on the filter wheel to properly attenuate UV contamination from the background and any bright UV/VIS sources, namely: the thin filter consisting of 160~nm of polyimide coated with 40~nm of aluminium (two of these were available on the FW), the medium filter consisting of 160~nm of polyimide coated with 80~nm of aluminium, and the thick filter consisting of 300~nm of polypropylene coated on one side with 100~nm of aluminium and on the other side with 100~nm of aluminium + 25~nm of tin. The RGS back illuminated CCDs, mounted in a row following the curvature of the grating Rowland circle, were equipped with an on-chip optical blocking filters consisting of a thin Al layer, ranging between 75~nm and 45~nm from the central to the outer chips, isolated from the Si by a \chem{MgF_2} isolation layer (about 26~nm thick)\cite{denherder2001}.

Contrary to the Chandra ACIS CCD, the EPIC cameras did not show any significant molecular contamination after the first five years of mission operation \cite{Kirsch2005}; some moderate decrease of the low energy response has been detected since the early 2010s on the MOS CCDs, the contamination was well modeled as a thin layer of pure carbon deposited on the surface of the cameras \cite{Smith2019}. No evidence of any contamination is detected on the pn-CCD. 

The \textbf{Neil Gehrels Swift} observatory \cite{Gehrels2004}, a NASA MIDEX mission dedicated to the study of gamma-ray bursts, was launched into a low Earth orbit (perigee 547~km, apogee 562~km, inclination \SI{21}{\degree}) in November 2004. The X-ray instrument uses a grazing incidence Wolter I telescope with 4.7~m focal length (spare flight module of the JET-X program) to focus X-rays onto an MOS CCD developed within the XMM-Newton program (EPIC camera). The CCD was protected by UV/VIS light by a single fixed optical blocking filter consisting of a 184~nm thick free-standing polyimide film coated on one side with 48.8~nm of aluminium. The XRT camera does not present significant evidence of molecular contamination affecting the low-energy response.

The Japanese ASTRO-E2 (\textbf{Suzaku}) X-ray observatory mission, launched into a lower Earth orbit on July 10, 2005 (perigee 550~km, apogee 550~km, inclination \SI{31}{\degree}) \cite{Mitsuda2007}, carried on board two soft X-ray instruments, the High-Resolution X-Ray Spectrometer (XRS), based on a small array of doped Si microcalorimeter detectors \cite{Kelley2007}, and the X-ray imaging Spectrometer (XIS) based on CCD detectors \cite{Koyama2007}. The XRS used a combination of five filters to protect the cryogenic detector from UV/VIS/IR radiation. The two innermost filters were free-standing polyimide 74~nm thick + 50~nm of aluminium, the two middle ones consisted of a free-standing polyimide film 100~nm thick + 100~nm of aluminium, and the outermost and largest one consisted of a polyimide film 100~nm thick + 80~nm of aluminium supported by a Ni mesh with 78\% open area. Unfortunately, the XRS has not gone into full operation due to a malfunction in the cryostat that caused a complete loss of the liquid He reservoir about 29 days after launch. 

The XIS on board Suzaku \cite{Koyama2007} had four X-ray MOS CCD cameras at the focal plane of four separate grazing-incidence reflective optics consisting of tightly nested, thin-foil conical mirror shells. Each camera was equipped with an optical blocking filter consisting of 140~nm polyimide coated on both sides with 80~nm and 40~nm of aluminium, respectively. The OBFs were located at a 2~cm distance from the CCDs and were kept approximately at \SI{-40}{\celsius}, the same temperature of the CCD. Shortly after the XIS door was opened, an unexpectedly high molecular contamination rate was recorded, different for each sensor and varying within the position of the OBF, being maximal at the center. The contaminants were identified to be predominantly C and O with a number ratio C/O $\approx$6 \cite{Koyama2007}. A numerical analysis was performed to identify the origin of the contamination. The model was compatible with the molecular contamination outgassing from the spacecraft’s materials, in particular: diethylhexyl phthalate (DEHP, \chem{C_{24}H_{38}O_4}), dibutyl phthalate (DBP, \chem{C_{16}H_{22}O_4}) or n-butyl benzyl phthalate (BBP, \chem{C_{19}H_{20}O_4}). Five years after launch, the total accumulated mass at the center of the OBFs was $\sim$\SI{300}{\micro\gram\per\centi\metre\squared}, nearly two times greater than that at the edges of the filters \cite{Urayama2012}.

The Japanese ASTRO-H high-energy observatory mission (renamed \textbf{Hitomi}) \cite{Takahashi2016} was launched on February 17, 2016, into a low Earth orbit (perigee 560~km, apogee 581~km, inclination \SI{31}{\degree}). Hitomi and its instruments performed extremely well up until the loss of communication with the observatory on March 26, 2016. One of the main instruments on board Hitomi was the Soft X-ray Spectrometer (SXS) based on an X-ray micro-calorimeter array with doped-silicon sensors \cite{Kelley2016}. The SXS instrument performed nominally on-orbit, satisfying all of its performance requirements. In particular, the array measured average energy resolution at 6~keV was 5~eV, significantly better than the expected 7~eV requirement, proving also that the filters had survived the launch without any measurable damage. The three outer filters consisted of a polyimide film 100~nm thick coated with 100~nm aluminium, supported by a 96\% open area Si mesh (diameters 35~mm, 24~mm, and 18.5~mm, respectively). The two inner OBF with a diameter of 12~mm consisted of a free-standing polyimide film 75~nm thick coated with 50~nm of aluminium \cite{DeVries2017,Kilbourne2018}.

The \textbf{eROSITA} X-ray telescope, the primary instrument on the Spectrum-Roentgen-Gamma (SRG) mission, was launched in July 2019 into the Lagrange point L2, about 1.5 million kilometers away from the Earth, and in December of the same year started to perform the all sky survey in the 0.2-8.0~keV \cite{Predehl2021}. The eROSITA instrument comprises an array of seven identical and co-aligned Wolter-I type mirror modules with 1.6~m focal length, each one consisting of 54 nested nickel shells coated with gold. The focal plane detectors are back illuminated pn-CCDs with $\sim$ 30 mm x 30 mm sensitive area covering a 1 x 1 square degree FOV \cite{Meidinger2020}. To protect the CCDs from UV/VIS light, a combination of an on-chip filter and a selectable filter from a filter wheel was implemented. An on-chip filter consisting of 200~nm of aluminium was deposited on the sensor chip of 5 out of 7 cameras. The cameras with on-chip aluminium use filters consisting of 200~nm thick free-standing polyimide, while cameras without the on-chip aluminium use filters with 200~nm thick free-standing polyimide + 100~nm of aluminium. External warm filters are needed also to protect the cold camera from molecular contamination. The detectors and optical blocking filters were not protected by a vacuum vessel during launch.
The two CCDs without on-chip aluminium (mirror modules 5 and 7) have been found to be contaminated by optical light whose intensity depends on the orientation of the telescope with respect to the Sun. The characteristics of the light leak are now better understood after the first all-sky survey coverages, and mitigation actions are being discussed to improve the low-energy response of these two cameras. At the current stage of the mission, there is no significant indication of in-orbit degradation in low-energy sensitivity due to molecular contamination.

\section{Functional goals}
\label{Functional_goal}

Filters on high-energy space experiments can either be in a fixed position in the beam path towards the sensitive detector or be mounted on movable systems, such as rotating filter wheels, to be selected on demand only for specific science observation cases.

Depending on the detector technology and science objective, filters can have one or more functional goals that would include:

\begin{itemize}
    \item rejection of out-of-band photons (UV, Vis, IR) which may increase the detector background, cause an energy calibration shift or, for some astrophysical sources (e.g. hot O-B stars), even overcome the x-ray fluxes;
    \item reduction of count rates for bright x-ray sources. Be filters or Al coated Kapton foils, a few tens of microns thick, are used when it is necessary to block the low energy photons and allow detection of higher energies (e.g. Fe-K complex). Neutral density etched metal foils with ~1\% transmission can be used when a flat attenuation is needed over the full spectrum.
    \item selection of a given wavelength band. Such filters can be coupled to photon counting detectors with pour energy resolution or to photometric detectors to perform spectroscopic analysis (e.g. observation of the solar corona).
    \item thermal control of a cryogenic detector by attenuating IR thermal radiation from warm surfaces in the solid angle visible from the sensitive detector area.  
    \item barrier to molecular contamination out-gassing from the spacecraft, especially in the first phases of operation in space.
    \item RF shielding from EMI generated by telemetry and on board electronics.
    \item shielding from low-energy (up to a few tens of keV) charged particles. At higher energies up to few hundreds of keV (e.g. soft protons), charged particles are not stopped by thin filters, however, they can be scattered and lose part of the energy \cite{Lotti2017}.
\end{itemize}

\section{Requirements and design drivers}
\label{Requirements}

Thin filters for X-ray detectors on board of space missions are usually identified as one of the critical items of the instrument, since their correct design and operation over the entire lifetime of the mission is essential for the success of the experiment. A partial failure of one or more filters in space can cause a reduced detector performance (e.g. increased threshold sensitivity, reduced efficiency, deteriorated energy resolution), or in the worst case even the non-operability of the detector (e.g. excess radiative load onto a cryogenic detector).

As the case for any subsystem operating in space, reliability is one of the key drivers in the design, and the goal to obtain optimal performance shall be less relevant than the adoption of a robust and already proven technology. Despite this general concern, it is meaningful and worth in the early phases of a mission development (A and B) to explore innovative solutions, which may provide a significant improvement in the experiment performance. In this phase, an iterative learning cycle is carried out consisting of design, technology breadboards procurement, characterization tests, and lessons learned. At the end of phase B (preliminary design), the adopted technology should be consolidated and only minor design changes should be acceptable in the early phase C (design consolidation).

During Phase-A (feasibility analysis), the design is mainly driven by the overall scientific requirements specified in the approved mission concept and is the result of a “bottom up” process, which takes benefit from the heritage of previous successful projects. In the following phases, a “top down” approach shall be followed to make sure to allocate to each subsystem a requirement value such that the sum meets the relevant top level requirement budget. The “bottom up” and “top down” processes shall converge into a reasonable allocation where the subsystem requirement values are achievable without a too large gap with respect to the early phase proposed design. Requirements usually evolve throughout the project development from a “high level” (Phases 0/A, “bottom-up” exercise) to more precise values in Phase B, and to frozen values at the end of Phase C (consolidated design and start of the procurement phase).

Requirements driving the design of a sub-system can be divided into few major categories. In the following we list the main requirement categories and make some specific examples for the design of filters:

\begin{itemize}
\item Scientific/Mission Performance Requirements e.g. X-ray transmission at specific energies in the sensitivity range of the instrument (this is the main science driver for the filters), level of out-of-band attenuation at specific wavelength ranges, materials, and impact onto the non X-ray background. The presence of pinholes may affect the out of band attenuation, thus a maximum number and size of pinholes is also usually specified.
\item General Requirements e.g. lifetime and acceptable aging effects, accessibility and interchangeability during the integration and verification phases, redundancy (not easy to implement for filters except for those mounted on a wheel), failure propagation.
\item Technical Requirements  e.g. thermal requirements (operating temperatures of the filters, considering that filter temperature may need to be regulated for contamination control, implying requirements on temperature monitoring and needed heating power), dimensions to account for full illumination of the detector, maximum applicable differential pressure, vibro-acoustic susceptibility, optical requirements (spatial uniformity of the filter to not affect the focal plane image quality), shielding requirements (filters may need to be conductive to close Faraday cages and guarantee proper RF attenuation values).
\item Environmental Requirements e.g. ambient conditions on ground, during launch and in space (temperature, humidity, cleanliness), thermal environment (radiative and conductive coupling with the environment and mounting structure, acceptable bake out temperature), mechanical environment (sine, random and shock vibration load during launch). In experiments where the filters are launched in atmospheric or residual pressure, requirements on acoustic loads during launch shall also be specified.
\item Interface Requirements e.g. positions, dimensions, mass and volume, mechanical interface to mount structure, alignment.
\item Operational Requirements e.g. thermal control during venting, launch, decontamination.
\item Calibration and Commissioning Requirements e.g. X-ray transmission accuracy
\item Development, Verification, Test, Ground Support Equipment Requirements e.g. safety check after integration and tests. 
\item Logistic Support Requirements e.g. packaging, transportation, handling, and storage. 
\item Quality Assurance Requirements e.g. cleanliness level of the rooms where filters are stored, characterized and integrated into the instrument.
\end{itemize}

\section{Materials and technologies}
\label{Mat_Tech}
Polypropylene (\chem{C_3H_6}), parylene (\chem{C_8H_8}) and polyethylene terephthalate (\chem{C_{10}H_8O_4}) have been used as structural support filter materials for some time since the early stage of X-ray space exploration. They have X-ray transmittance and bandpass performance similar to carbon, but they are generally stronger than  graphite. Polypropylene with $\sim$\SI{1}{\micro\metre} thickness was typically obtained by stretching a film a few tens of micrometers thick. Parylene films have been manufactured with thicknesses as low as 100~nm, in addition, parylene has higher temperature resistance with respect to polypropylene, and has thus been preferred for solar observation experiments \cite{Spiller1990}.

In the 1980s, Lexan (\chem{C_{16}H_{14}O_3}, bisphenol-A polycarbonate) became the polymer of choice because it proved much stronger than materials previously used. Aluminium coated Lexan filters have been employed on EXOSAT \cite{Taylor1981}, ROSAT \cite{Truemper1982}, EUVE \cite{Malina1992}, and Yohkoh \cite{Tsuneta1991}. Thin polycarbonate films are produced by means of spin coating techniques.  The film thickness is controlled by spin coating parameters, solution viscosity, and curing temperature. The aluminum is evaporated in more runs, sometimes distributed over more days.

\begin{figure}[htb]
    \centering
    \includegraphics[height=6cm,width=8cm]{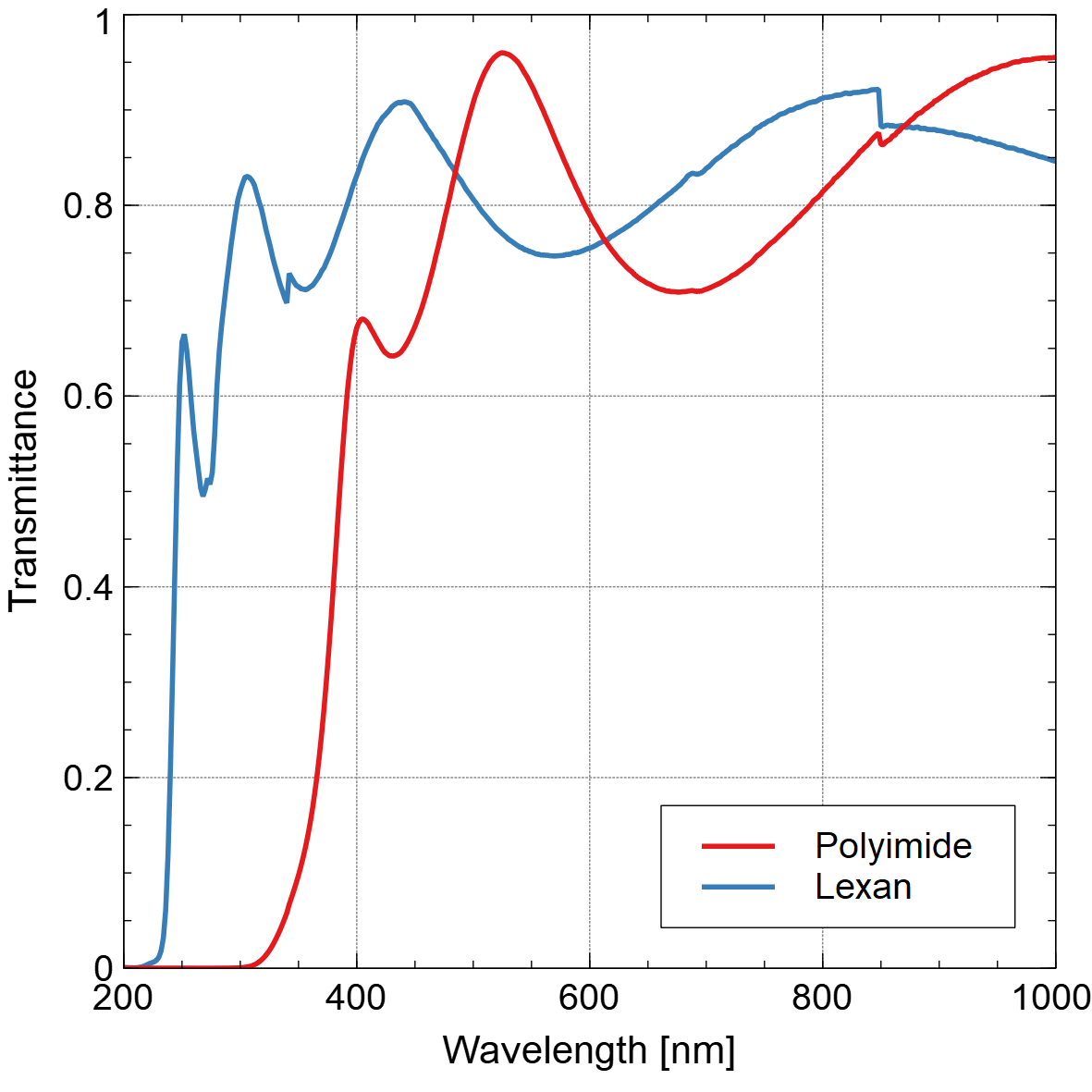}
    \caption{Measured UV/VIS transmission vs. wavelengths for a sample of luxfilm polyimide 274.5~nm thick (red) and a sample of Lexan 281.6~nm thick (light blue) in the wavelength range 190-1000~nm.}
    \label{fig:mater}
\end{figure}

Within the development program of the UV/Ion shields of the Chandra High-Resolution Camera (HRC) microchannel plate detector \cite{Zombeck1995,Murray1997}, an investigation was carried out on different plastic materials that could provide high UV rejection concurrently with high transmittance in the soft X-ray region of interest (E~$>$~100~eV). A comparison was performed between Lexan, luxfilm polyimide, \chem{C_{22}H_{10}N_2O_5}), Parylene-C (\chem{C_8H_7Cl}), which has one hydrogen atom in the aryl ring of parylene replaced by chlorine, and VYNS (\chem{C_2H_3Cl}, poly(vinyl chloride)) \cite{Barbera1994}. VYNS and parylene-C were ruled out as an alternative to Lexan being also transparent in the UV. Polyimide turned out to be a possible alternative, being more opaque than Lexan in the UV. Figure \ref{fig:mater} shows a comparison of UV/VIS transmission performed on two samples of bare luxfilm polyimide and Lexan filters, nearly 280 nm thick, within the development activities of the Chandra HRC. Polyimide provides a significant attenuation in the wavelength range 200-300~nm where the Lexan film is already fully transparent.

Further investigations performed on luxfilm polyimide by LUXEL showed this polymer having superior strength with respect to Lexan, which allows for thin windows to withstand some differential pressure, and having more ductility that allows for even thinner filters and windows that will survive space mission environments while improving mission throughput. Polyimide is also stable across a wider temperature range than previously used polymeric filter materials \cite{Powell1997}. 
Investigating the mechanical properties of polyimide thin films at very low temperature is a difficult task, preliminary results of an investigation performed by LUXEL shows that polyimide thin films ($\sim$500~nm thick) keep their yield strength (as indicated by measured burst pressure) down to 4~K, in addition, polyimide mechanical properties can be improved by optimizing the production process and in particular increasing the curing temperature \cite{Grove1999}. More effort is, however, needed to perform a more extensive characterization of the mechanical performances of thin films at very low temperatures and to optimize the curing process for very thin films ($<$~100~nm).

Upon these encouraging results, since mid 1990's polyimide has become the most used material for filters to operate high energy photon detectors in space. Missions for solar exploration that have employed polyimide filters include: the Solar and Heliospheric Observatory (SOHO) launched in 1995, the Advanced Composition Explore (ACE) launched in 1997, the Transition Region and Coronal Explorer (TRACE) launched in 1998, SOLAR B (Hinode) launched in 2006, and the Solar Dynamic Observatori (SDO) launched in 2010. High-energy Space observatories using polyimide filters include: Chandra launched in 1999, XMM-Newton launched in 1999, Neil Gehrels SWIFT observatory launched in 2004, ASTRO-E2 (Suzaku) launched in 2005.

Cross metal wire grids, plated metal meshes or photolithographic etched meshes have been developed to support low leak soft X-ray proportional counter windows capable to sustain atmospheric gas pressure, and to mechanically support large area thin filters capable to resist to the vibro-acoustic load of a launch into space \cite{Powell1997}.

Photolithographic Si fine meshes \cite{Nelms2002} have also been developed as mechanical support structure for thin filters. They have been employed on a few experiments using X-ray microcalorimeters such as the X-ray Quantum Calorimeter Sounding Rocket Program, a joint effort between the University of Wisconsin and Goddard Space Flight Center \cite{McCammon2008}, the Micro-X sounding rocket experiment, using a transition-edge-sensor X-ray micro-calorimeters \cite{Wikus2010}, and the SXS experiment on board Hitomi \cite{Kilbourne2018}. The Si mesh is obtained by Deep Reactive Ion Etching of a Si wafer, and thanks to the good thermal conductivity of Si, heaters mounted on the outer circular frame allow to warm-up the filters for contamination removal. The Si meshes are quite brittle and thus they cannot be rigidly tightened to a metal supporting ring. For this reason, they are attached on a kinematic mount which has the disadvantage of not being leak tight, thus requiring the use of a moisture shield to minimize contamination of the inner shields.

Photolithographic polyimide meshes have been developed and used in space on one of the plasma protection windows of the LECS experiment on board Beppo SAX \cite{Parmar1997}. More recently, PI meshes have been investigated further by LUXEL for applications where good thermal conductance of the filter and low frequency RF attenuation is not a concern \cite{Grove2010}. PI meshes are interesting since they become fully transparent at photon energies above a few kiloelectronvolts.

Silicon nitride windows have been initially investigated as low leak rate windows for soft X-ray proportional counters \cite{Aucoin1994}, and very thin ($\sim$40~nm) membranes with a diameter up to 31~mm supported by a poly-Si mesh with $\sim$75\% open area have been manufactured proving to be leak tight and capable to withstand up to three bars of differential pressure \cite{Torma2013,Torma2014}. The same authors have started an investigation to manufacture larger diameter windows ($\sim$100~mm) of thin silicon nitride membranes supported by Si mesh suitable as thermal filters for cryogenic microcalorimeters. Present results do not show any significant advantage of this technology with respect to the more consolidated polyimide technology when large size filters need to be manufactured. 

Thin films of carbon nanotubes (CNT) are investigated since a few years as a novel UV transparent electronic material with excellent and tunable electrical, optical and mechanical properties \cite{Wu2004}. Very large size (hundreds of square centimeters) and very thin (a few tens of nanometers) free standing CNT pellicles have been manufactured with very high EUV transmittance, good mechanical strength, and thermal stability in vacuum \cite{Timmermans2018}, such material has been recently considered also as a promising material for large size X-ray filters and is currently being investigated within a research contract funded by the European Space Agency and lead by AmeteK Finland Oy.

Another potentially interesting material for the realization of thin high performing X-ray windows is graphene, which has low atomic number, gas tightness, high chemical stability, high electrical conductivity, light blocking ability \cite{Nair2008}, nontoxicity, and is one of the strongest material ever measured \cite{Lee2008}. Small size windows of this material (diameters $<$ 10~mm) have been manufactured with thicknesses of the order of \SI{1}{\micro\metre} as an alternative to Be windows for energy dispersive X-ray detectors \cite{Huebner2016}.

To conclude this section, we mention the main commercial actors with some heritage in manufacturing thin filters for space applications. 

HS-Foils (Finland), recently acquired by AMETEK, is specialized in silicon nitride windows, free standing or mesh supported to withstand high differential pressure. The company is involved in research activities on innovative CNT based filter materials for space applications.

Lebow (California, USA) supplies a large variety of foils of different materials (not all made in house), either free-standing or supported by electroformed Ni or Cu mesh, and etched stainless steel mesh.  Filter materials include several elements and compounds and a few type of polymers. 

Luxel (Washington, USA), manufactures ultra-thin filters for space applications since the early seventies and is still active being the main actor worldwide in the space sector (see the web site www.luxel.com for a detailed description of contributions to a long list of space programs). Luxel provides some standard filters, also for ground laboratory applications (e.g. SEM, TEM, high energy physics), although its specialty is in custom filter design. 

Materion (Ohio, USA) is a large company specialized in innovative material for research and industrial application. Specialized in beryllium filters and optics, has participated to different space missions including James Webb Space Telescope.

Moxtek (Utah, USA) is a high tech company focused not only on filters. They produce ultra-thin windows including polymers. They have manufactured filters for some space applications including the thin and medium filters of the EPIC camera on board XMM-Newton.

\section{Performance modeling}
\label{Performance_modeling}
Filter design, as discussed in section \ref{Requirements}, requires a delicate trade-off between several conflicting needs, and relies on accurate performance modeling to reach an optimal compromise. The X-ray transparency requirement, calling for thin membranes, is in direct contrast to achieving out-of-band shielding and mechanical robustness, and thermal properties play an important role too, especially when the filters are to be integrated into a cryogenic instrument.

\subsection{X-ray transmission}
X-ray propagation through a solid is subject to absorption by the photoelectric effect, incoherent (Compton) scattering, coherent (Rayleigh) scattering, and Bragg diffraction.  Other photon-matter effects, such as electron-positron pair production and photo-nuclear effect, are negligible for energies of the photon $<1$~MeV. \citet{Hubbell1969} gives a comprehensive review about the main processes involved in the interaction between X-rays and condensed matter.

\begin{figure}[htb]
    \centering
    \includegraphics[height=8cm]{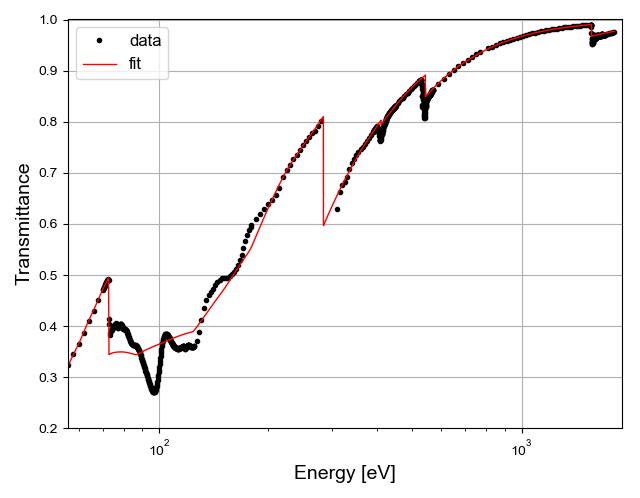}
    \caption{X-ray transmittance of a filter made of polyimide (nominal thickness 45~nm) and aluminium (nominal thickness 30~nm). The black dots are a measurement performed at a synchrotron beamline. Data in the carbon edge is not reported in the plot due to contamination that compromises the data reliability in that region. The solid red line is a transmittance model that does not take in account the fine absorption structures in the edge regions.}
    \label{fig:x-transmittance}
\end{figure}

The photoelectric effect is the primary contributor to the absorption of X-ray radiation in a thin film up to energies of about 30~keV. An X-ray photon can interact with an electron pertaining to an atomic inner shell, being completely absorbed and transferring its energy to the electron, which is therefore removed from its orbital. However, if the photon energy is lower than the binding energy of the electron, the process cannot take place and the photon cannot be absorbed by this mechanism. For this reasons, X-ray transmittance in function of the photon energy shows characteristic absorption edges in proximity of the binding energies of the shells (see fig. \ref{fig:x-transmittance}). The photoelectric effect probability is strongly dependent on $Z$, the material atomic number, and $E$, the photon energy, varying several order of magnitudes for different materials and energies. In particular, the probability due to interaction with the K-shells is proportional to $Z^5/E^{3.5}$ \cite{Heitler1954}.

Incoherent scattering (also Compton or inelastic scattering) takes place when a photon interacts with a free electron or weakly bound electron in an outer shell, knocking it out from the electron cloud and losing some energy in the process. The transmission loss due to incoherent scattering is negligible, with respect to photoelectric effect, for photon energies up to 30~keV, except for interaction with very low Z materials where a small contribution cannot be neglected above a few kiloelectronvolts.
Coherent scattering, also known as Rayleigh or elastic scattering, is a process in which a photon interacts with an atom as a whole (not only with one electron) and it is scattered substantially maintaining all of its energy. It is coherent in the sense that the photon energy does not change while its propagation direction and phase can change, thus resulting in interference effects. If the material is highly ordered, as in crystalline materials, and has a certain thickness, the coherent scattering leads to Bragg diffraction. For X-ray filters Bragg diffraction is often not taken in account, since they are usually thin and/or composed of amorphous or disordered materials; thick filters, like the beryllium one used as instrument gate valve on Hitomi, are bound to show Bragg features in the transmittance spectrum \cite{Eckart2018}. 

The “narrow beam” transmittance of a filter is defined as $T=I/I_0$, where $I_0$ is the impinging X-ray radiation intensity and $I$ the intensity of the radiation exiting from the filter in the same direction of $I_0$.  The concept of a narrow beam is related to the fact that the radiation scattered around, even if it crosses the filter, is not considered as part of $I$. The contribution of each illustrated effect can be taken in account through the linear attenuation coefficient, a quantity related to the probability of the photon-matter interaction event to occur. If $\mu$ (cm$^{-1}$) is the sum of the linear attenuation coefficients of the aforementioned effects, the transmittance of a homogeneous medium can be written as:
$$T=e^{-\frac{\mu}{\rho} \cdot \rho t},$$
where $\rho$ is the material density (g cm$^{-3}$) and $t$ its thickness (cm). X-ray filter materials are usually composed by several elements; in this case, the transmittance is given by:  

$$T=e^{-\sum_i{(\frac{\mu}{\rho})_i \cdot \rho_i t_i}},$$
having indicated with $\mu_i$, $\rho_i$ and $t_i$ the quantities relative to the i-th element. The same equation holds if the filter is a multi-layer since, with a good approximation, only the total number of atoms for the various elements in the photon path is relevant, not how they are arranged in the layers.
The quantity $\mu/\rho$ (cm$^2$ g$^{-1}$) is called mass attenuation coefficient, and is tabulated for many elements and materials. For an extensive review on the experimental and theoretical available data about mass attenuation coefficients see \citet{Hubbell2006}. \citet{Saloman1988} gives tables of experimental attenuation data in terms of the cross-section $\sigma$ (in units of barns/atom, other authors give it in cm$^2$/atom), where $\sigma=\frac{\mu}{N}$ ($N$ is the number of atoms per volume unit), for element Z=1 to Z=92 and energies from 100~eV to 100~keV. See also \citet{Henke1993} and \citet{Hubbell1975} (erratum in 1977). The available tabulated data is generally valid when it is possible to assume that the atoms scatter independently and there is no influence by the chemical structure of the material. This can be considered true for energies sufficiently far from the absorption edges and $>\sim100$~eV. Closer to the edges, the absorption presents a finely structured curve, related to multiple-scattering and resonance processes and a dependency to the local atomic structure surrounding the atom that is interacting with the photon. These absorption structures, for which theory and modeling can be found in \cite{Stern1974, Rehr2000, Zabinsky1995, Koningsberger1987}, have a significant impact when a detailed knowledge of a filter spectral response is required. High energy-resolution transmittance measurements can be performed to obtain an accurate calibration (see sect. \ref{Calibration}).

\subsection{UV/VIS/IR transmission}
\label{IR-VIS-UV transmission}
Rejection of radiation with wavelength from the IR to the UV is crucial to block light from astrophysical sources, stray-light from bright sources that are not directly in the field of view, usually the Sun, the Earth or the Moon, and thermal IR radiation originated in the surfaces of the instrument with a light path to the detector. The last point is especially relevant for cryogenic detectors, such as microcalorimeters, operating at $T<100$~mK, which are affected by noise generated from thermal fluctuations due to incoming radiation.

Filters can block such unwanted radiation by absorption or reflection. When leveraging the latter, it is important to take in account multiple reflections that can bring back rejected light, eventually allowing it to reach the detector. Scattering through the filter medium can usually be neglected or assimilated in a limited way to transmission or rejection, as far as the detector is concerned.

Transmittance and reflectance of a multi-layer filter are affected by the interference of the waves determined by the filter structure. For a multi-layer composed of isotropic and homogeneous films, they can be calculated from the knowledge of the complex refractive index and the thickness of each film by a matrix method based on Fresnel equations, very well suitable for computer evaluation \cite{Heavens1960}. \citet{Palik1997, Hagemann1975} have published tables including optical constants for many materials, and several online databases allow to readily obtain refractive index data from a multitude of literature sources \cite{refractiveindex.info, filmetrics.com, pvlighthouse, sopra}. Nonetheless, many aspects related to the manufacturing process can have an impact on the refractive index, and a characterization of the specific employed materials may be needed. 

\begin{figure}[htb]
    \centering
    \includegraphics[height=8cm]{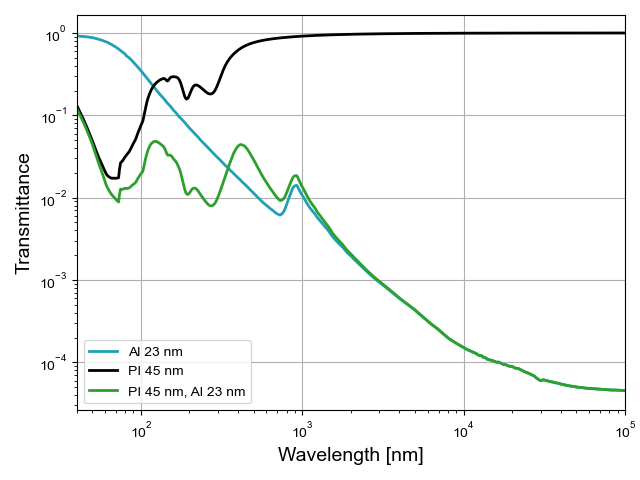}
    \caption{UV/VIS/IR transmittance models of an aluminium film (23~nm), a polyimide film (45 nm) and a combination of the two above layers of aluminium and polyimide. The curves are obtained using the modeling described in this section.}
    \label{fig:uv-vis-ir-model}
\end{figure}

One of the current state-of-the-art standard filter for astrophysical X-ray detectors is a multi-layer consisting in a polymeric (e.g. polyimide) membrane coated with a metal (e.g. Al). We can consider, as a relevant example, a multi-layer with four films: PI, aluminium, and two aluminium oxide layers, one on each face of the Al. The aluminium oxidation is a drawback, since it is quite transparent to the unwanted radiation, but it absorbs x-rays the same way aluminium does. Al provides a good rejection by reflection for wavelengths higher than about 100~nm, and gives an absorption contribution between 4~nm and 1000~nm, particularly effective between 4~nm and 17~nm, with an inter-band absorption peak at about 830~nm. PI gives an important contribution absorbing UV radiation in the range 8~nm - 400~nm. Transmittance curves in fig. \ref{fig:uv-vis-ir-model}, obtained using the modeling here discussed, show the described behavior. The presence of pin-holes in a filter can reduce the rejection performances. The Al deposition process is especially susceptible to particle contamination that can result in pin-holes in the metal layer. The issue can be mitigated by coating the polymer on both sides, but this solution has its drawbacks: the total amount of aluminium oxide is doubled, the sandwich structure can worsen the degradation of the polymer due to atomic oxygen trapping \cite{Banks2004} and, depending on the polymer thickness, interference effects can reduce rejection in specific bands.

Optical constants for pure metals are in general well known and tabulated in literature, but the deposition process, especially when dealing with thin films, can affect significantly the interaction with light. The specific process used to deposit the metal, typically e-beam or thermal evaporation deposition or sputtering, together with parameters such as the substrate temperature during deposition, have an impact on the film density and continuity. Discontinuities of the film, or a lower density, reduce the light rejection of the film.

Polymers have largely varying optical properties based on their formulation. \citet{Hasegawa2001} gives a detailed review on the photo-correlated properties of polyimides. Light absorption takes place when the energy of the incoming photon matches the gap between two allowed energy states \cite{Manivannan1997}. This can occur at UV wavelength with electronic excitation, where an electron can move from a bound orbital to an unbound one. $\pi$ bonds of the most common groups give place to absorption in the range 160~nm - 400~nm, while $\sigma$ bond absorption peaks are usually at lower wavelengths. In the IR, the absorption is determined by allowed molecular vibration modes, which strongly depend on the structure of the molecule (bond lengths, bond angles, and geometry). Forouhi-Bloomer dispersion formulation for amorphous materials \cite{Forouhi1986, Forouhi2019} can well describe the refractive index of polyimide in the whole IR - extreme UV range. A multi-oscillator Lorentz’s model can be used to describe the refractive index in the IR \cite{Zhang1998, Kawka2001}.

Different techniques are used for characterizing the optical properties. A prism coupler can be used to measure the refractive index at specific wavelengths \cite{Lee2003, Ando2002}. Reflectance spectroscopy measurements at different angles and polarizations, performed on multiple PI samples with different thicknesses, have been used to derive the refractive index in the range 5~$\mu$m - 1.4~$\mu$m \cite{Kawka2001}, and a similar technique has been employed to evaluate the thickness and the refractive index for a specific wavelength in the IR \cite{Lvesque1994}. Normal incidence IR transmission spectroscopy has been used by \citet{Zhang1998} to determine the thickness of several PI samples and their refractive index. Other thin film characterization techniques are reported in \cite{Heavens1960, Frey2015}.

\subsection{Mechanical and thermal analysis}
\label{Mech_thermal}
Thin filters for space missions are subject to severe mechanical and thermal stresses: launch vibrations and acoustic load, differential pressures, and changes from room to cryogenic temperature. Increasing mechanical robustness implies thicker films and stronger supporting meshes, thus a reduction in the transparency to X-rays. 

Finite element method (FEM) simulations help optimizing the design, to guarantee the necessary robustness while minimizing the opacity; nonetheless, the quality of the materials has to be experimentally tested (as discussed in section \ref{Characterization}) since defects and inhomogeneities can easily dominate filter failure. FEM simulations are employed to study single layer freestanding membranes \cite{Small1992}, bi-layers, and failure mechanisms \cite{Li2011}. 

Bulge test, displacement measurement of a membrane subject to a differential pressure, has been commonly adopted to retrieve experimentally the mechanical properties required for the simulation of freestanding membranes \cite{Vlassak1992, Bhandarkar2016}. 

Regarding vibrations, the membrane is generally so thin that it can be neglected when simulating a structural mesh mechanical response by FEM. Similarly, in terms of thermal behavior, the mesh is the main contributor to the conduction along the filter plane, and a meshless filter is usually a poor conductor and can result in relevant temperature gradients along its surface.

The thermal emissivity $\epsilon$ of a membrane, useful for modeling the thermal interaction of the filter with the surrounding environment, can be retrieved as a function of its wavelength $\lambda$ by the relation $\epsilon(\lambda) = 1-T(\lambda)-R(\lambda)$, where $T$ and $R$ are the transmittance and reflectance discussed in section \ref{IR-VIS-UV transmission}.

\section{Characterization techniques}
\label{Characterization}

The ultimate performance of the filters in space, until the end of life of the mission, shall be the result of careful design activity in response to the instrument requirements, a detailed plan of characterization and qualification activities, and a well understood, controlled, and repeatable manufacturing process.
Heritage from previous missions, knowledge of the ongoing investigations on new promising technologies, and performance modeling, described in the previous section, provide the main inputs to choose the baseline filter technology and identify a preliminary design. 
Optical and mechanical characterization tests, performed on partially representative filter breadboards, shall then be performed in phase A to verify the feasibility of the investigated technology and to prove that, with a proper design, filters can meet the main scientific requirements (X-ray transmission, out of band attenuation). Results of the preliminary tests will guide the choice of the filter materials, the type of supporting structure if needed, the frame material and mounting scheme, the type of bonding between the thin filter, the supporting structure, and the frames. 
In phase B, when the system interface requirements start to be consolidated, characterization tests shall be performed on more representative models to assess the maturity of the technology and the compliance with the main technical and environment requirements (e.g. static and dynamic mechanical loads). 
In the following, we will briefly describe a few techniques used to characterize filter breadboards. Some of these techniques are also used for the qualification and acceptance tests of more advanced models, including the flight units.

    \subsection{X-ray transmission spectroscopy and imaging}

During the design, development, and production phases of filters for any X-ray mission, several experimental measurement campaigns must be conducted both to support the design for improving the X-ray transmission and to calibrate the response of the instrument, taking into account the presence of one or more filters that attenuate incident radiation \cite{Kohmura2000}\cite{Puccio2020}. 
A good design of the filters has to ensure the maximum X-ray transmission, in particular at low energies (E $<$ 1 keV) while retaining the required out-of-band rejection, thickness uniformity, and mechanical robustness. Furthermore, X-ray transmission spectroscopy and imaging are also essential during the calibration phases (see section \ref{Calibration}) \cite{Chartas1996} \cite{Eckart2018}.

The main goals of X-ray transmission spectroscopy and imaging measurements of the filters include:
\begin{itemize}
    \item verification of the compliance with the requirements (see sect \ref{Requirements});
    \item development of a refined model to predict filter response by deriving optical constants from reference samples (including edge regions);
    \item derive the areal densities ($\rho_it_i$) and, knowing the densities, the thicknesses ($t_i$) of the different material layers of the filter, and their spatial homogeneity; 
    \item identification of suitable measurement facilities and achievable accuracy for the flight filters calibration;
    \item calibration of flight filter witness samples (or directly flight filters if feasible).
\end{itemize}

Experimental campaigns can be carried out using collimated monochromatic beams available in synchrotron X-ray beamlines such as FCM \cite{Gottwald2006} and SX700 \cite{Sokolov2014} at Bessy II, BEAR at Elettra \cite{Nannarone2004}, and Metrologie at Soleil \cite{Idir2010}, or by use of a wide beam produced by electron impact X-ray sources in laboratory based facilities such as XACT \cite{Barbera2006}, Panter \cite{Friedrich1998} \cite{Freyberg2006}, or Beatrix \cite{Salmaso2021}.

\begin{figure}[htb]
    \centering
    \includegraphics[height=5.5cm]{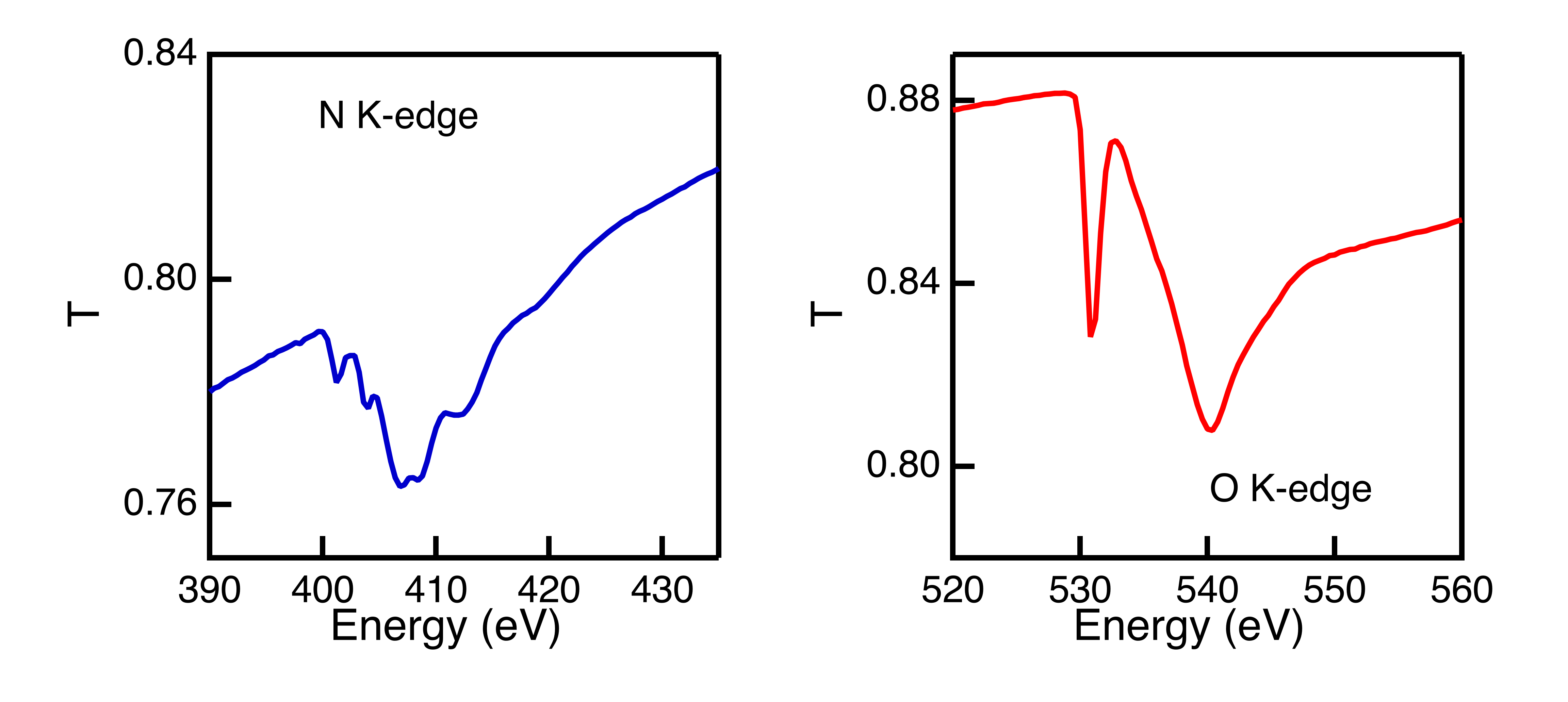}
    \caption{X-ray absorption edges on a filter made of polyimide (45~nm) and aluminium (30~nm) measured using synchrotron radiation. Left panel: N K-edge. Right panel: O K-edge.}
    \label{fig:edge}
\end{figure}

A synchrotron beamline experimental set-up may collect spectra in a wide energy range (figure \ref{fig:x-transmittance}), with a high spectral resolution and accuracy that is specifically required for the absorption edge regions (figure \ref{fig:edge}). 

In the absorption edges, filter transmission changes rapidly as a function of energy, therefore, in order to get few \% accuracy on the instrument quantum efficiency the x-ray transmission in the edges shall be calibrated with a small energy step, typically smaller than the required absolute energy scale calibration accuracy. As an example, in the case of the X-IFU instrument on Athena, the 1 sigma energy scale calibration accuracy is currently set to 0.4 eV, therefore, the x-ray transmission of filters inside the edges will be calibrated with a step $<$ 0.2 eV. Different beamlines show distinctive characteristics; hence collecting and comparing data recorded in two or more beamlines is a functional choice \cite{Puccio2020}. 
Synchrotrons produce high brilliance X-ray radiation, many orders of magnitude greater than any laboratory source. This property allows the use of high spectral resolution monochromators collecting a reasonable transmitted flux in energy regions where samples may be opaque to the radiation, obtaining low noise and statistically sound measurements. Spatially resolved transmission measurements of filters at specific energies can give information on the uniformity of the membrane thickness, which is a fundamental technical requirement as reported in section \ref{Requirements}. These characteristics combined with the high stability of these sources make synchrotrons an essential tool for filter investigation and calibration. Notwithstanding the high accuracy of these measurements, there are a few disadvantages with this kind of source. The experimental chambers available at synchrotron beamlines do not often offer enough room to host large samples as the flight and spare flight models are, thus collecting the transmission spectra of large filters is quite challenging, especially for the imaging since a 2D-motorized stage is also required. Some facilities, designed also with an eye on calibration activities, offer this valuable option as is the case for the PTB beamlines at BESSY II \cite{Gottwald2006,Sokolov2014}.  Furthermore, the high brilliance might damage filters. Such an occurrence can be mitigated by moving the samples out-of-focus to reduce the energy density and minimize any damage and any chemical or structural modification due to the beam.

The laboratory based facilities (e.g. XACT, Panter, Beatrix, see the article "Facilities for X-ray Optics Calibration" in this handbook for more details) are typically equipped with large experimental chambers, able to accommodate large filters, whole filter stacks or even fully assembled instruments. Moreover, by employing different techniques, they produce wide collimated beams (tens of centimeters in diameter), allowing to perform X-ray transmission maps of large area filters when coupled with an imaging detector such as a CCD or a MCP detector. For comparison, the typical size of a synchrotron beam in the main experimental chamber does not usually exceed a few square millimeters making the wide area x-ray mapping more complex and time-consuming. On the other hand, the spectrum generated by a laboratory beamline is not white, but it features emission lines characteristic of the source anode and a less intense Bremsstrahlung continuum. The much lower brilliance with respect to synchrotron radiation makes it difficult to operate with a very narrow bandwidth monochromator (that has a very low efficiency), and the measurements are usually performed only in the emission lines without the possibility to achieve the energy resolution of the synchrotrons.
    
    \subsection{UV/VIS/IR spectroscopy}

As reported in section \ref{Requirements}, the rejection of the undesired out-of-band radiation is a critical requirement for any X-ray detector. For instance, semiconductor detectors such as the DEPFET active pixel arrays must be shielded from UV/VIS photons to prevent energy scale offsets, and microcalorimeters must be shielded by IR radiation to avoid a spectral resolution deterioration by photon shot noise.

The out-of-band radiation can be blocked using filters manufactured with appropriate materials (sect.\ref{Mat_Tech}) and proper design (sect. \ref{Requirements} \cite{Barbera2016}). As an example, the baseline design for the large optical filter of the WFI detector of Athena is a bilayer of polyimide and aluminium with thicknesses 150~nm and 30~nm, respectively \cite{Barbera2018b}.

For the abovementioned reasons, the characterization of both UV/VIS/IR transmission and reflection plays a key role in the development and the design of optical blocking and thermal filters. In this respect, planning and performing UV/VIS/IR measurement campaigns \textit{ex-situ} on sets of witness samples is essential to choose suitable materials and determine the thickness of each selected layer. 

\begin{figure}[htb]
    \centering
    \includegraphics[height=5.5cm]{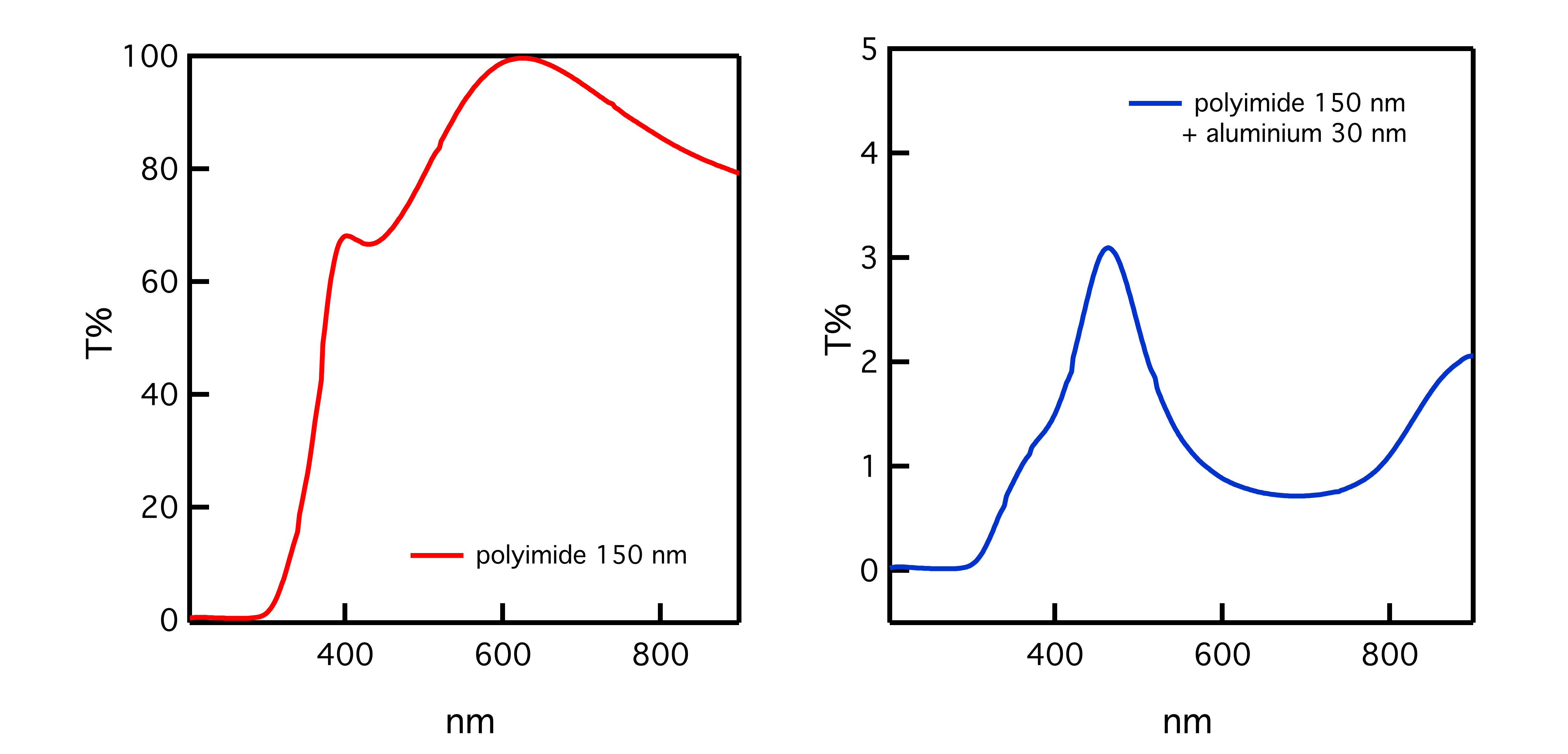}
    \caption{Ultraviolet and Visible transmission of  a  150~nm membrane of polyimide (left panel), and  a  150~nm membrane of polyimide coated with 30~nm aluminium (right panel).}
    \label{fig:T_UV_VIS}
\end{figure}

The transmission of a 150~nm membrane of polyimide and a 150~nm membrane of polyimide coated with 30~nm aluminium in the UV/VIS spectral range is reported in fig. \ref{fig:T_UV_VIS}.

For the reported case, the thickness of the polymer guarantees low transmission in the UV region (below 300~nm), and the aluminium attenuates efficiently the visible radiation because of its high reflectivity in this wavelength range. 

Optical spectra in the ultraviolet, visible, and near-infrared regions can be acquired using a double beam spectrophotometer while the infrared spectra in the near-, medium-, and far-infrared are usually recorded using a Fourier Transform spectrophotometer. Furthermore, the thickness uniformity over a large area of a prototype sample can be verified by mapping the UV/VIS/IR transmission/reflection in several points. 

The experimental UV/VIS/IR campaigns allow also the retrieval of the actual refractive index of a specific material that was produced with a certain manufacturing process, such information is essential to develop a proper transmission model as reported in section \ref{Performance_modeling}. 

Another important parameter that can be derived from IR measurements is the emissivity already mentioned in section \ref{Mech_thermal}. In fact, the manufacturing process can affect deeply the emissivity of a material since it can modify considerably the surface roughness. The knowledge of the emissivity of the filter surfaces is very relevant especially for detectors operating at low temperatures which are very sensitive to thermal radiation from any surface in their field of view, including the filters.

It is well-known that the sum of transmissivity $\tau$, reflectivity $\rho$, and emissivity $\epsilon$, for a certain wavelength $\lambda$, must be equal to 1,  $\tau(\lambda) +\rho(\lambda)+ \epsilon(\lambda) =1$. The $\epsilon$ can thus be calculated from the experimental measurement of $\tau$ and $\rho$ as a function of wavelength.

    \subsection{X-ray photoelectron spectroscopy}

Materials can be subject to ageing and be chemically altered over time when stored in presence of air \cite{Kohmura2000}; hence the requirements reported in section \ref{Requirements} might be met in a pristine filter but might not after the component has been in storage waiting for integration in the instrument. Such modifications often involve the outer layers, which can be investigated by X-ray photoelectron spectroscopy (XPS), a technique that only probes the outermost atomic layers. The information collected is restricted to a depth of few nanometers (1~nm - 10~nm).

\begin{figure}[htb]
    \centering
    \includegraphics[height=6cm]{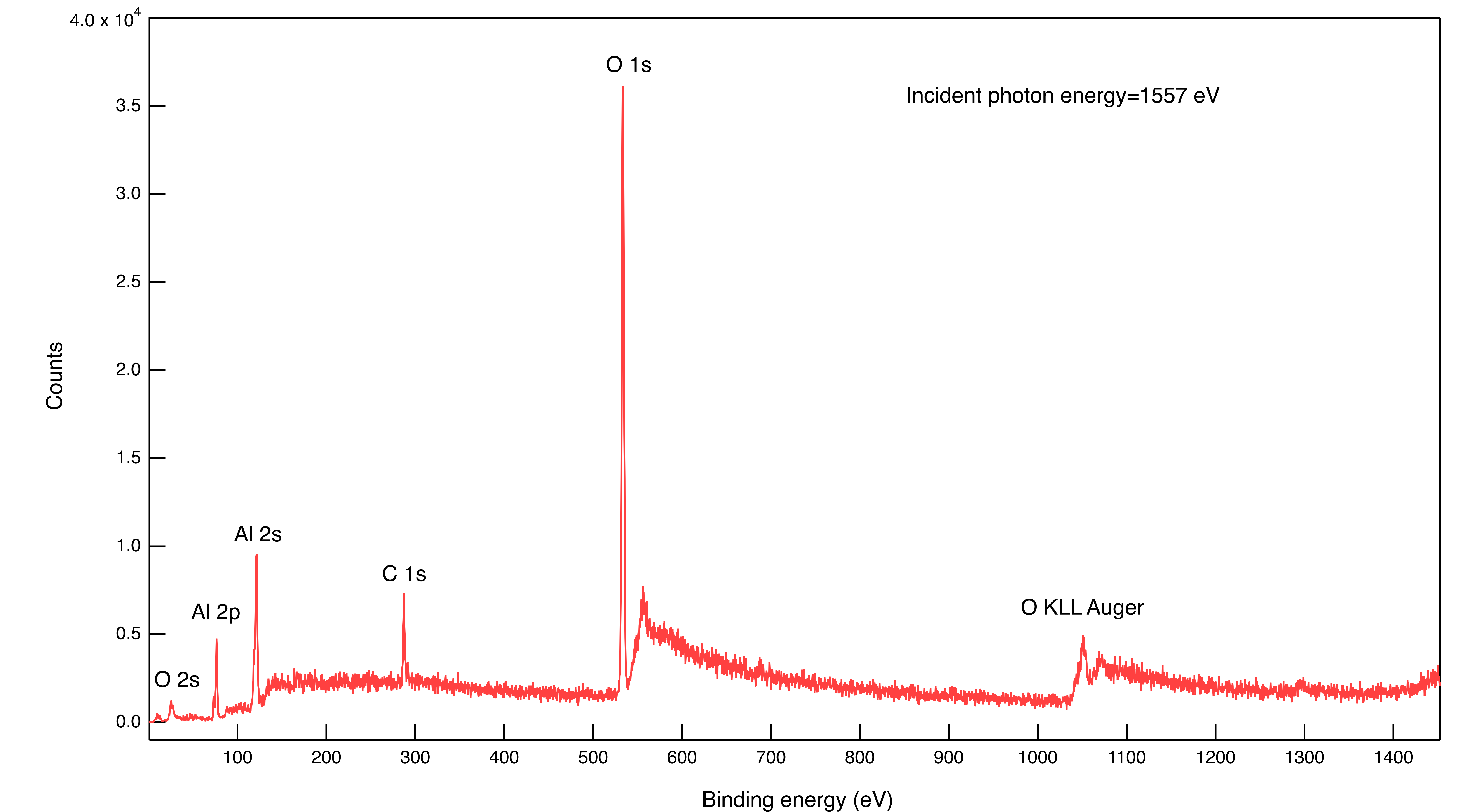}
    \caption{XPS survey recorded at 1557~eV incident photon energy. In the plot the O 2s, Al 2p, Al 2s, C 1s, C 1s, O 1s, the O KLL Auger peaks are identified.}
    \label{fig:xps}
\end{figure}

The XPS is based on the photoelectric effect; if an X-ray photon with appropriate energy hits a material, the ejection of an electron, often named photoelectron, can occur. The kinetic energy of the ejected photoelectron, originating from a specific atomic shell, can be measured. The simple relevant relation is reported in eq. \ref{eq:xps_1} where $E_i$ is the energy of the photon energy, $E_k$ and $E_b$ are the kinetic energy and the binding energy, respectively, of the outgoing photoelectron, and $w_f$ is the work function that is typical of the employed instrument \cite{Stevie2020}.
\begin{equation}
\centering
    E_i=E_k+E_b+w_f
    \label{eq:xps_1}
\end{equation}
The knowledge of $E_i$ and the accurate measurement of $E_k$, via Eq. \ref{eq:xps_1} allows to recover the $E_b$ of all the ejected electrons from a specific orbital of the present chemical elements but hydrogen or helium. The $E_b$ contains information about the chemical element, its oxidation state, and its chemical environment \cite{Stevie2020}.
An example of XPS wide scan, called survey, of a sample is reported in fig. \ref{fig:xps}

XPS spectra can be acquired using two kinds of X-ray sources: an electron impact source, usually using an aluminium or magnesium anode, or synchrotron radiation, a highly tunable and brilliant source.

The analysis of a single peak may be useful to compute the thickness of an overlayer onto a bulk material \cite{Jablonski2019}. For instance, the filters for X-ray detectors are usually coated with metallic aluminium, which rejects out-of-band radiation, that is naturally oxidized \cite{Barbera1997}. This chemical change of the outer layers may affect the spectroscopic performances of the filter, causing a deterioration of the detector performances. The XPS analysis allows to measure the calculation of the oxide thickness, thence helping to properly design the thickness of each layer of the flight filters \cite{Sciortino2016}.
    
    \subsection{Radiofrequency shielding effectiveness}
    Highly sensitive X-ray detectors can be hindered by electromagnetic noise originating from the spacecraft telecommunication system or from on board electronics. In such cases, Faraday cages are built to shield the sensor, and filters are required to act as cage extensions, providing a suitable attenuation to radio frequency (RF) radiation. Filters composed of a thick metal foil are able to block most of the RF, but thin films designed to maximize X-ray transparency may be unable to reflect the radiation entirely, allowing a fraction of the power to pass through; metal meshes with a proper pitch can help to increase the shielding effectiveness (SE) of such filters.

\begin{figure}[htb]
    \centering
    \includegraphics[height=8cm]{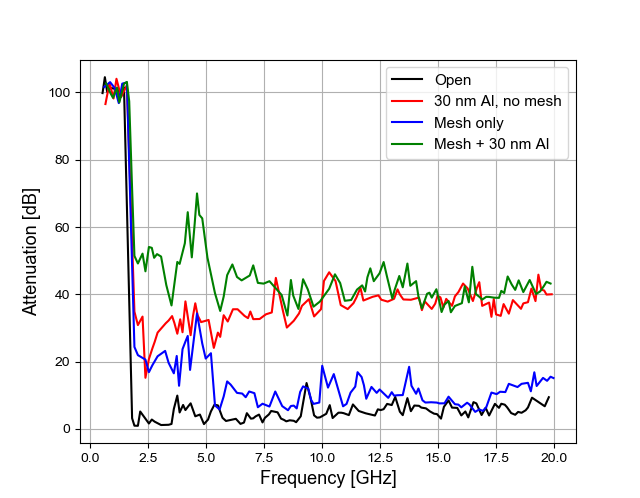}
    \caption{Attenuation measured in a reverberation chamber with a diameter of 100~mm \cite{LoCicero2018}. The cut-off at 1.76~GHz is due to the geometry of the chamber. During the measurement, the filter under test is mounted on a metal wall in the middle of the chamber, with a 56~mm diameter hole. The black curve refers to the measurement without any filter (the attenuation of the holed wall is measured), and the others refer to samples consisting of a 30~nm thick Al foil, a 4~mm pitch copper mesh, a combination of mesh + Al foil.}
    \label{fig:rf-attenuation}
\end{figure}

Measurements conducted on thin film filters in a reverberation chamber have shown that Al films of 30~nm are able to provide an attenuation of about 30~dB for frequencies above 5~GHz, while metal meshes with 4~mm pitch can provide an attenuation of about 15~dB for frequencies below 5~GHz \cite{LoCicero2018} (see Fig. \ref{fig:rf-attenuation}).

When performing RF shielding measurements, the geometry and the RF properties of the filter operating-environment has to be taken into account; usually metal or metallized walls are defining the space where the RF can propagate inside satellite instruments, which can be seen as an electrically large enclosure. Such an environment can support a multitude of superimposed electromagnetic modes, each one potentially having resonance peaks inside the cavity. The geometry of the enclosure is determinant to the total SE, a small aperture has a lower cut-off frequency for supported electromagnetic modes, and its size may be sufficient to ensure high attenuation for low-frequency radiation, even without a filter. 

A free-space propagation attenuation measurement can be misleading, since it would just emulate a specific field configuration, unlikely to be supported into an enclosure, not taking into account resonances that can result in a strongly reduced SE for the corresponding frequencies. Foreseeing how external radiation is coupling with the enclosure and which modes would propagate is impractical, so the SE characterization is performed statistically with a great number of different field configurations. 

\begin{figure}[htb]
    \centering
    \includegraphics[height=5cm]{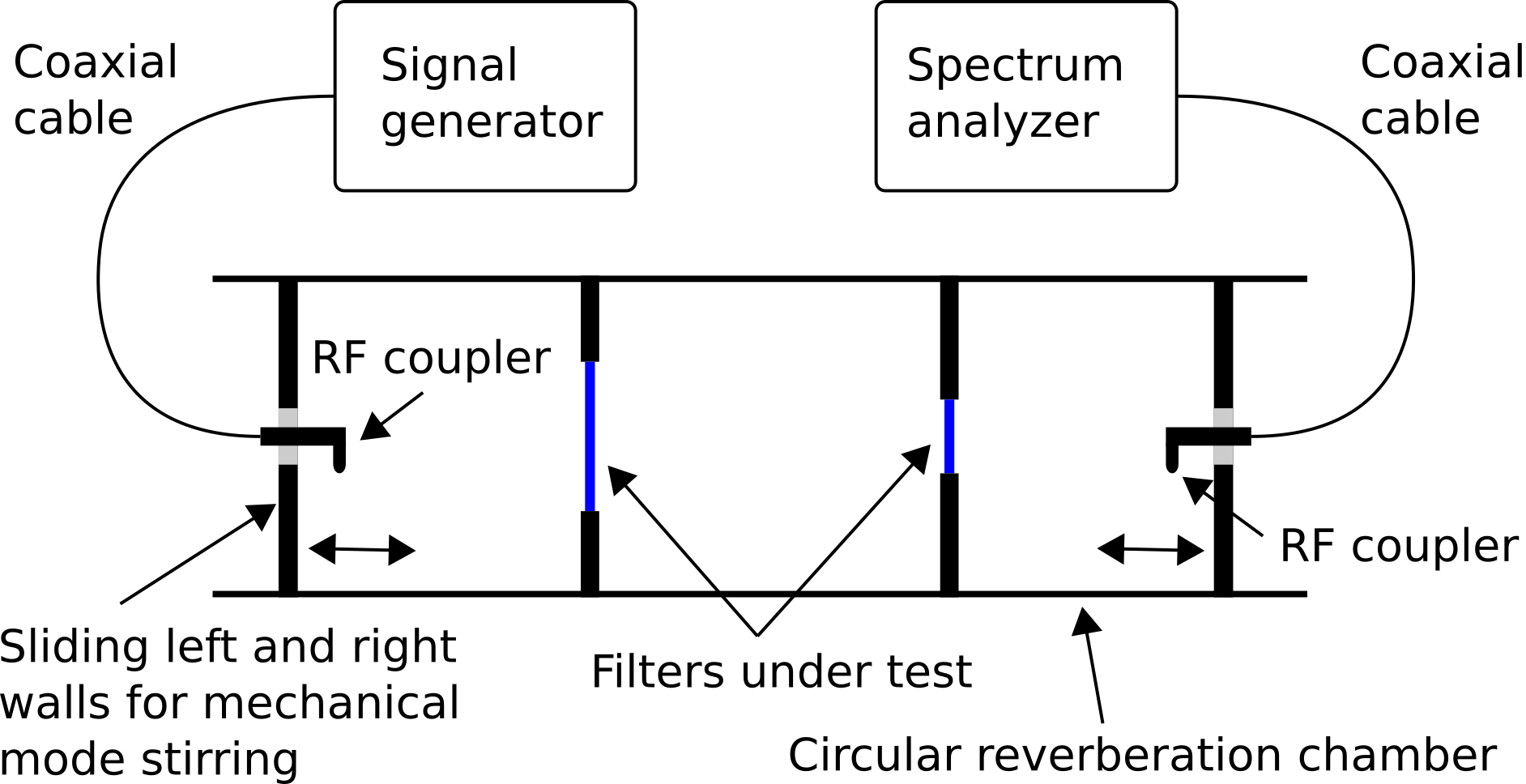}
    \caption{Schematic drawing of a set-up to measure the shielding effectiveness of two filters, using a reverberation chamber.}
    \label{fig:reverberation-chamber}
\end{figure}

A reverberation chamber, an enclosure with high reflectivity and geometry representative of the filter operative environment, allows to obtain the needed variety of configurations using mechanical or frequency mode-stirring \cite{Serra2017, Hill1994}. It is equipped with two RF couplers, usually antennas or loops, facing each other across the chamber. One or more filters are placed inside the enclosure in a representative geometrical configuration. The measurement is performed by a spectrum analyzer or a network analyzer, injecting an excitation signal through one coupler and measuring the signal coupled to the second one. The chosen mode stirring technique must then be applied to acquire the data needed for statistically calculating the SE in the frequency range of interest. See figure \ref{fig:reverberation-chamber} for a schematic of the measurement setup. Care must be taken in properly characterizing the measurement system, performing measurements in the following conditions: open (without filters), closed (filter aperture closed with a metal slab), and dark (signal measured without excitation).
    
    \subsection{Imaging and microscopy}

\begin{figure}[htb]
    \centering
    \includegraphics[height=3.8 cm]{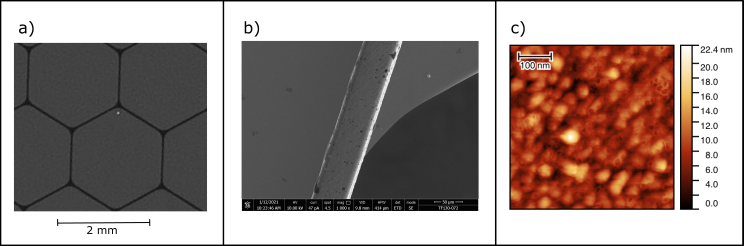}
    \caption{Images collected at different scales: a) Image acquired using a high-resolution photographic scanner to detect coarse defects such as pinholes, b) SEM image of a supporting mesh wire, and c) AFM image of a portion of aluminium coating onto a filter membrane.}
    \label{fig:imaging}
\end{figure}

The collection of images at different magnification levels (surveying dimensions from nanometers up to millimeters) is a powerful tool to check integrity and quality of filters in any step of the qualification, including acceptance tests after the delivery. An example of different collected images is reported in fig. \ref{fig:imaging}

The integrity of the filters must be verified after the production and after any test. A high-resolution photographic scanner, or a digital camera, allows freezing information on the structural integrity of both the membrane and the mesh (fig. \ref{fig:imaging}a). In addition, these tools enable to look for defects in the membrane or in the gluing between membrane and mesh. The search for pinholes in the membrane and its coating  or contaminating particles  (typical diameter \SI{5}{\micro\metre}) can be performed using an optical microscope with the appropriate resolution, equipped with a digital camera and a motorized translator to automate the image acquisition. These measurements can be performed in properly clean environments also on representative models such as the flight and flight spare units.
Optical and thermal filters are assembled with methods reported in section \ref{Mat_Tech}, often adopting metal coatings. The roughness of these coatings can adversely affect the rejection of out-of-band radiation. The quality of the coating must be checked in terms of waviness (millimeters scale), roughness (few nanometers), and the presence of pin-holes or other imperfections.
The investigation of small-scale defects, including changes in roughness or non-uniformity at the nanometric level, requires the use of more sophisticated techniques such as scanning electron microscopy (SEM) (fig \ref{fig:imaging}b) and atomic force microscopy (AFM) (fig \ref{fig:imaging}c).
These last characterizations can be performed using witness samples of the flight filters, which are fully representative and manufactured in the same production batch with the same processes. The SEM images allow to assess the quality of the supporting mesh bars (surface morphology, roughness, cleanliness of the cuts, walls verticality...) or membrane to mesh bonding (integrity and uniformity)\cite{Townsend2016}. Additionally, the AFM images allow inspecting the morphology of the membrane and metallic coating surfaces extracting a z-profile to compute the surface roughness\cite{Deoliveira2012}.
    
    \subsection{Environmental tests}
The instrumentation on Space missions suffers the impact of high-energy particles in the energy range from few keV to hundreds of MeV coming from the Sun and other energetic astrophysical sources. Soft protons of solar origin may be scattered by X-ray grazing incidence optics and be partially focused onto the focal plane with potential risk to damage the detector and/or increase its background noise; this effect was identified for the first time on Chandra \cite{Prigozhin2000} and XMM-Newton \cite{Turner2001,Struder2001} observatories, posing severe new constraints on the design and operation of upcoming X-ray missions with larger effective areas of the mirrors. 
Filters, used to protect the X-ray detectors in Space, may partially absorb, scatter, and modify the energy spectrum of soft protons, thus playing a fundamental role in the determination of this detector background component \cite{Lotti2017}.
Particles with sufficient energy, typically cosmic rays with E~$>$~100~MeV, can cross the spacecraft and reach the focal plane from every direction. The collisions with such energetic particles can seriously damage filters producing an increase of the surface roughness or degradation of mechanical, chemical, and optical properties \cite{Li2007} \cite{Kim2008}.

Irradiation tests are conducted on filters at proton accelerator facilities, with particle energies in of few megaelectronvolts, to evaluate aging effects. Typical qualification doses for a few years mission in Space are of the order of 10$^{10}$~\si{\per\centi\metre\squared} 1~MeV equivalent. Surface analysis of the filters by AFM, XPS, and X-ray transmission measurements are performed before and after the irradiation to establish the occurrence of aging effects.

Our solar system is filled with micrometeoroids (natural mineral grains likely originated from asteoroids and comets) and debris produced by artificial objects placed in orbit. As known from previous X-ray missions such as XMM-Newton and Swift \cite{Carpenter2006}, and the eROSITA German X-ray satellite launched in July 2019 and presently operating in the Lagrangian point L2 \cite{Meidinger2021}, micro-meteoroids can pass or scatter through the grazing incidence mirror systems or be fragmented there and reach the focal plane procuring local damages to the sensitive area of the detectors. The particle fluxes decrease with increasing particle size in the range \SI{0.1} - \SI{100}{\micro\metre}, have speeds in a typical range 1-20~\si{\kilo\metre\per\second}, and particle densities ranging from \SI{1}{\gram\per\centi\metre\cubed} to \SI{5}{\gram\per\centi\metre\cubed}  \cite{Carpenter2006} \cite{Perinati2017}. 

No clear evidence has been reported so far of damages on the thin filters placed along the focused X-ray beam optical path of X-ray missions. Severe damages, consisting of a complete disruption of a filter, would have been easily recorded, while minor damages, consisting of small size pinholes, might have occurred and not been detected because they are not affecting the detector performances measurably. Despite this, a great concern on the behavior of thin filters under micrometeoroids bombardment is put in the design of future X-ray observatories with large collecting area telescopes like Athena, where the expected number of micrometeoroid hits at the focal plane is quite significant. Given the predicted particle flux in L2 reported in \cite{Rodgers2015}, and the Athena telescope open area, it is possible to estimate a particle fluence at the telescope focal plane of a few tens per year of \SI{0.1}{\micro\metre} size, few per year of \SI{1}{\micro\metre} size, and an unlikely occurrence of \SI{100}{\micro\metre} size particles over the lifetime of the mission. 

As reported in \cite{Perinati2017}, a collision of a particle on a filter can result, as a function of its mass, density and relative speed, on either: 
\begin{enumerate}
    \item the particle is stopped by the filter;
    \item the particle punctures the filter, passing through and leaving a hole;
    \item the particle is fragmented or evaporated upon impact, eventually passing through the filter as a shower of smaller particles.
\end{enumerate}

Specific experimental tests will be performed on breadboard filters of the Athena focal plane detectors at a dust accelerator facility capable to accelerate particles with Space representative sizes, speeds, and compositions. This type of tests allow to evaluate the effects of micrometeoroids impact on the filters and their role in partially protecting the detector.

Missions flying in low Earth orbits (LEO) face the issue of atomic oxygen (AO) corrosion \cite{Reddy1995}. Diatomic oxygen and ozone in the upper atmosphere are subject to photo-dissociation by solar radiation, leading to a high concentration of extremely reactive atomic oxygen. It is well known, since the first shuttle missions, that polymeric materials are highly susceptible to deterioration when exposed to AO. Strategies to overcome the issue span from the design of AO resistant polymers to protection with a variety of thin film coatings. Filters made of metalized polymer have a inherent protection from the AO, provided that the metal film is facing the external environment and that its quality is good enough. Micro defects in the metal coating constitute entrance windows for the AO, leading to the corrosion of the polymer. Such imperfections can be more difficult to avoid when the metal film thickness has to be kept within a few tens of nanometers. An additional coating can be helpful, particularly high-quality extra-thin layers deposited with high performance processes such as atomic layer deposition \cite{Cooper2008}. Several methods have been used to test materials resistance to AO corrosion, many of which rely on chambers where oxygen plasma is generated by microwaves, laser pulses, or other techniques. To increase the relevance of the test, the kinetic energy involved in the material-atom interaction has to be taken into account, which in orbit is related to the speed of the spacecraft flying in the AO rich environment (a few kilometers per second). Testing setups able to bombard a sample with high-speed oxygen atoms have been built (e.g. ESTEC atomic oxygen simulation facility) to allow for accurate quantitative measurements of AO degradation effects.
    
    Stress to filters can be originated not only from direct mechanical solicitations (vibrations, acoustic loads) but also from variations of temperature or humidity. Operating temperature for X-ray filters spans from as low as a few tens of mK, when dealing with cryogenic detectors, to about 340~K, for decontamination procedures. Moreover, they can be subject to a large number of thermal cycles, when their temperature is not regulated, because of the intermittent exposure of the spacecraft to the Sun. Such temperature variations produce thermo-mechanical stresses, potentially inducing failures during the mission lifespan.

Thermo-vacuum tests are performed to check filters behaviour when subjected to multiple thermal cycles in high-vacuum, making use of temperature controlled chambers connected to vacuum pumps. Temperature control can be attained with mechanical coolers or cryo-coolers, coupled with electrical heaters. Applied temperature ramps are usually in the order of a few K/min. 

Humidity can affect surfaces chemically (e.g. water-soluble materials, corrosion), but it can as well induce mechanical stresses when materials are hygroscopic; this is the case of polyimide, which absorbs water from the atmosphere, swelling considerably. The majority of the water content is quickly released when the PI is put in high vacuum, giving place to a contraction of the membrane. During space operation, humidity uptake is not a concern, but flight filters are usually subject to humidity during integration in the instrument, and may be subject to humidity cycles if repeatedly brought from air pressure to vacuum during instrument testing. Depending on the materials of choice, humidity stress tests can be performed into a climatic chamber (a temperature and humidity controlled chamber) or into a vacuum vessel (cycling between vacuum and ambient pressure).

    \subsection{Mechanical loads}
A spacecraft and its instruments must successfully go through mechanical tests before it is allowed to be flown. In particular, it has to survive without any measurable degradation the severe environment of a launch. The sources of vibro-acoustic excitation are mainly due to noise generated by the launcher during ignition, lift-off and atmospheric flight. In addition, since the fairing evacuates in a few tens of seconds while the launcher exits the atmosphere, it is also very important to properly test the venting procedures to make sure that differential pressure does not build up onto sensitive items as it is the case for thin fragile filters. Differential pressure onto filters can also build up during handling, storing and testing in laboratory, for this reason proper procedures shall be prepared and followed to protect the filters during all phases of the development. 

Vibro-acoustic qualification tests of filters are performed during the development phases to prove the validity of the design, to assess the maturity of the investigated technology and to verify the validity of mathematical modeling used to search for the natural frequencies and to evaluate the stress regime under severe launch environment. In the later phases of the program, when representative models have to be delivered to the leading agency or prime industrial contractor to perform functional, structural and thermal verifications or when the flight unit is ready for integration into the spacecraft, lower “acceptance” level vibro-acoustic tests shall be performed. 

The acoustic environment inside the fairing at launch is usually the primary source of vibration, in addition, other sources may contribute to the overall spectrum including possible torsional oscillation imparted by the launch vehicle onto the interface adapter, random vibrations, and sustained oscillations produced by the motors.

\begin{figure}[htb]
    \centering
    \includegraphics[height=6cm]{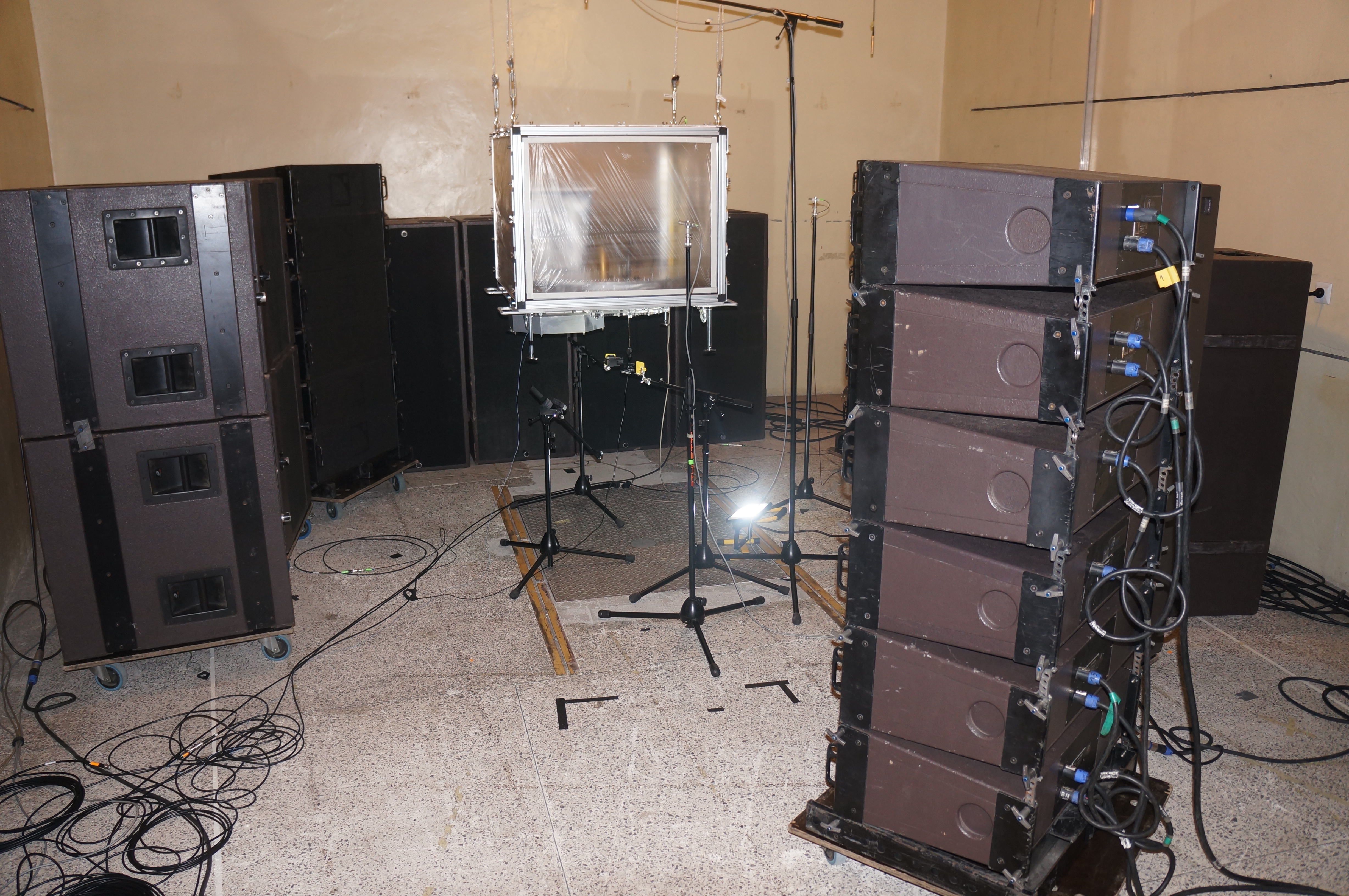}
    \caption{Acoustic test performed on a set of representative filters of the WFI detector on Athena inside a filter wheel breadboard. The FW is closed inside a plastic tent for cleanliness protection. Four microphones are standing up around the FW to control and monitor  the sound spectrum.}
    \label{fig:Acoustic tests}
\end{figure}

Acoustic tests are conducted in reverberant chambers providing a uniform sound field. The sound pressure is measured in different locations around the item under test by separate control and monitor microphones. Accelerometers can also be mounted on the system to measure the induced vibrations. Figure \ref{fig:Acoustic tests} shows the the reverberation chamber of the AGH University in Krakow during an acoustic test performed on a set of filters developed for the Athena mission. The filters were mounted inside a filter wheel breadboard covered by a plastic tent for cleanliness protection. 

The qualification acoustic spectrum depends on the specific launcher. The envelope spectrum of the noise induced inside the fairing of the Ariane 6 launcher, used in the above mentioned acoustic test of Athena filters, is shown in table \ref{tab:acoustic_noise}. It corresponds to a space-averaged level within the volume allocated to the spacecraft inside the fairing. 

\begin{table}[htb]
\begin{center}
\caption{Acoustic noise spectrum under the fairing of the Ariane 6 launcher (ref. Ariane 6 User’s Manual, Issue 1, Revision 0, March 2018)}
\begin{tabular}{|c|c|}
\hline
 \textbf{Octave Center Frequency}   & \textbf{Flight Limit Level}\\
\textbf{[Hz]}                       & \textbf{[dB]}        \\             
                                    &  (reference: $0\;dB = 2 x 10^{-5} Pa$)\\
\hline
\hline
31.5    & 128\\
\hline
63      & 131\\
\hline
125     & 136\\
\hline
250     & 133\\
\hline
500     & 129\\
\hline
1000    & 123\\
\hline
2000    & 116\\
\hline
\textbf{OASPL* (20-2828 Hz)}& 139.5\\
\hline
\multicolumn{2}{l}{\footnotesize *OASPL – Overall Acoustic Sound Pressure Level}
\end{tabular}
\label{tab:acoustic_noise}
\end{center}
\end{table}

Vibration qualification tests usually consist of three characteristic modes: 'sine' testing by progressive sweep of frequencies and amplitudes, 'random' testing applies random frequencies and amplitudes, and 'shock' testing induces a sudden severe excitation, simulating the shocks felt during stage separations and engine firings. Sine, random and shock tests on small mass items are performed by use of electrodynamic shakers.
When vibration tests are performed at sub-system level, the vibration mask to apply to the filters is derived by modeling the transfer function from the launch vehicle adapter to the sub-system under test. Vibration tests can be performed with filters in vacuum or in atmospheric pressure depending on the actual condition during launch. The use of a laser scanner vibrometer during a vibration test allows to measure displacements,  velocities, and accelerations along the filter surface with an high spatial resolution, and to derive natural modes of vibration. Vibration tests are performed at least on two orthogonal directions, one off-plane the filter membrane (the most severe) and one in plane, with the same levels applied for the two axis.

\begin{figure}[htb]
    \centering
    \includegraphics[height=4cm]{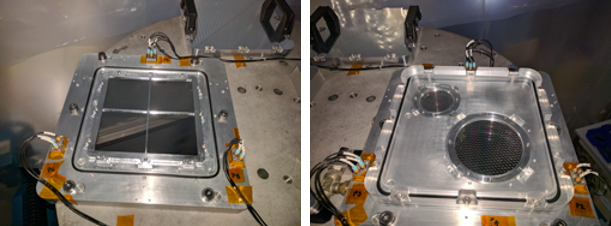}
    \caption{Filter breadboards of the Athena WFI (left panel) and X-IFU (right panel) detectors mounted in off-plane configuration on the shaker head at CSL in Liege.}
    \label{fig:shaker}
\end{figure}

Figure \ref{fig:shaker} shows a set of filter breadboards developed for the Athena WFI  (left panel) and X-IFU (right panel) detectors mounted on the shaker head at CSL in Liege to perform a qualification vibration test off-plane. 

Table \ref{tab:vibration_load} shows the sine, random, and shock reference vibration levels of the Ariane 6 launcher used in these tests as input to the transfer function derived from the FEM modeling of the whole Athena WFI instrument. 

\begin{table}[htbp]
\begin{center}
\caption{Sine, random, and shock reference vibration levels of the Ariane 6 launcher.}
\begin{tabular}{|c|c|}
\multicolumn{2}{c}{Sine (25.0 g 0-peak, sweep rate=2 oct/min)}\\
\hline
\textbf{Frequency range (Hz)}   & \textbf{Level}\\
\hline
5.0 - 23.0                      & 11.7 mm (0-peak)\\
23.0 - 100.0                    & 25.0 g (0-peak)\\
\hline
\multicolumn{2}{c}{}\\
\multicolumn{2}{c}{Random (16.9 g RMS, duration=150 s)}\\
\hline
\textbf{Frequency range (Hz)}   & \textbf{PSD}\\
\hline
20.0 - 100.0                    & +3.00 dB/oct\\
100.0 - 300.0                   & 0.5 $g^2$/Hz\\
300.0 - 2000.0                  & -5.00 dB/oct\\
\hline
\multicolumn{2}{c}{}\\
\multicolumn{2}{c}{Shock test reference level (axial), Q = 10}\\
\hline
\textbf{Frequency range (Hz)}   & \textbf{SRS(g)}\\
\hline
100                             & 20\\
1000                            & 400\\
3000                            & 400\\
\hline
\end{tabular}
\label{tab:vibration_load}
\end{center}
\end{table}

Very thin filters for X-ray detectors in space can only withstand limited differential pressure (as low as few millibars) before they go into plastic deformation and eventually break, for this reason, during the technology assessment phase it is also necessary to perform static pressure load tests. A bulge test consists in loading a free standing membrane with a differential pressure and measuring its profile and the maximum displacement as a function of the applied pressure \cite{Vlassak1992}.
    
\section{Calibration}
\label{Calibration}

Any calibration strategy of X-ray missions combines several measurements and analyses and involves calibration at component level, at sub-system level, at instrument level, in-flight by internal calibration sources, and in-flight by sky sources \cite{ODell1998}. 

While it is often impractical to bring a fully assembled instrument in front of a synchrotron beam for on-ground calibration, the use of bright and highly monochromatic synchrotron beams are commonly used for calibrations at component and sub-system levels (filters, detectors, mirror modules). Before planning a specific filter calibration, an accurate knowledge of the selected synchrotron beamlines must be achieved, consequently, a proper calibration strategy includes a fraction of the allocated time dedicated to well characterize the beam (e.g. intensity, spectral resolution, energy scale) and to evaluate measurement uncertainties, including systematics (e.g. higher-order X-ray contamination).
The calibration of filters must include both the broadband transmittance, the edge fine structure of the chemical elements that make up the filters (C, O, N, Si, Al,…), the filling fraction of the fine support mesh (if present), and the effects on transmission of any spilling adhesive used to bond the mesh to the thin filter (if used). The broadband transmission can be acquired using a coarse energy step (5-50~eV), while the edge fine structures require a finer energy step (0.1-0.3~eV) \cite{Eckart2018}. The edge region must be resolved with an energy step significantly smaller than the expected resolution of the instrument, and at least a factor three smaller than the required energy calibration accuracy.  Moreover, the transmission uniformity is checked performing measurements in several locations to derive an upper limit of the non-uniformity at different spatial scales. The blocking factor due to presence of a supporting mesh must be also measured. This task is sometimes not trivial when using X-ray collimated beams with sub-millimiter size. Depending on the chemical elements and the thickness of the meshes, these can be totally opaque at low energy, and semi opaque, or even transparent, at high-energy, as in the case of silicon meshes employed in Hitomi\cite{Kilbourne2018}. In the case the meshes are made up of high-Z metals (Cu, Nb, Ni, stainless steel, Au, Ag…), the mesh can be totally opaque as in the case of the Athena current baseline \cite{Barbera2018a}.

When more filters are present in a stack, as it is the case for the  X-IFU (Athena) and SXS (Hitomi) microcalorimeter detectors, calibration measurements are performed on each individual filter to maintain the flexibility to replace one or more filters if needed. Furthermore, performing measurements on the whole stack, though in principle useful to improve the measurements accuracy, is usually difficult due to limitations in the available room in the synchrotron measurement chambers. For instance, measurements of the full set of five SXS optical blocking filters were performed as consistency checks of the individual calibration measurements \cite{Eckart2018}.

A calibration plan for filters for X-ray missions may be sorted out in different phases as following \cite{Pajot2019}:
\begin{itemize}
\item calibration of standards. This consists in a set of experimental campaigns to obtain high quality data to recover the optical constants of the materials assembled to fabricate the filters. These high-resolution measurements must include the absorption edges of all the chemical elements present in the filters in the spectral range of interest. The edges must be acquired using standards with equivalent thicknesses of the selected design for the flight filters;
\item calibration of witness samples. In this phase, the transmission is measured on small filter samples manufactured from the same coated membrane and thus fully representative of the filter models under calibration. The measurements performed on the  witness samples allow to avoid handling flight and flight spare filters in critical environments.  If the previous phase is carried out properly, only the broadband transmission need to be collected without high spectral resolution measurements of the absorption edges.
\item Calibration of flight filters, prior to integration, and flight spare filters. The calibration of flight filters allows to validate the previous phase, to check the uniformity over a large area of the membrane, and to measure the filling factor of the meshes. 
\end{itemize}

\section{Future perspectives}
\label{Future}
The very successful records on the use of polyimide/Al filters reported by different X-ray space observatories launched since the late nineties, and in some cases still operating like Chandra and XMM-Newton, encourage to keep this material choice as the baseline for the design of new experiments on Space high-energy astrophysics observatories. At the same time, future missions equipped with larger diameter telescopes will need larger size focal plane filters with even improved X-ray transparency, especially in the soft band (E~$<$~1~keV), still providing sufficient of out of band attenuation and resistance to vibro-acoustic loads during launch. To achieve these goals the existing filter technologies have to be pushed to their limits, and highly X-ray transparent fine meshes will need to be implemented as supporting structures and cutoff attenuators in the RF \cite{LoCicero2018} or FIR \cite{Eckart2019}.  A few examples of pushing the current technology for filters are given below in the current design of the Athena and Lynx missions. 

The Advanced Telescope for High-Energy Astrophysics (Athena) \cite{Barcons2015} is the second Large (L2) astrophysics space mission selected by ESA in the Cosmic Vision 2015-2025 Science Program to address the Hot and Energetic Universe science theme \cite{Nandra2013}. Athena, whose launch is currently scheduled in the early thirties, will be equipped with a $\sim$2.5~m diameter, 12~m focal length grazing incidence X-ray telescope based on the innovative silicon pore optics technology \cite{Bavdaz2010}, delivering  $\sim$1.4~m$^2$ effective area at 1~keV with an angular resolution full width at half maximum (FWHM) of $\sim$8"  over a large field of view ($>$~40’ diameter) \cite{Collon2019}. The telescope will be mounted on a movable platform which will allow both focus adjustment and tilt to point the X-ray beam on one of the two focal plane detectors: the X-ray Integral Field Unit (X-IFU) \citep{Barret2018} microcalorimeter array, and the large Wide Field Imager (WFI) \cite{Meidinger2017} depleted p-channel field effect transistors (DEPFET) array.

The X-IFU array of microcalorimeters will operate at a temperature $<$~100~mK inside a sophisticated multi-stages detector cooling system (DCS). To allow the X-ray photons, focused by the large area Athena telescope, to reach the X-IFU detector at the focal plane, windows have to be opened on the cryostat thermal and structural shields. Thin and highly X-ray transparent filters  (named thermal filters, THFs) need to be mounted on such shields to attenuate the radiative heat load from warm surfaces. The currently investigated design consists of a stack of five filters operating at 50~mK, 2~K, 30~K, 100~K, and 300~K, respectively. In order to maximize the low energy response,  each filter will consist of a polyimide film 45~nm thick coated with 30~nm of aluminium, bonded to a thick metal mesh with $\sim$2\% blocking factor (BF) which provides mechanical support, thermal conductance, and RF attenuation at low frequencies  ($<$~6~GHz). The diameters of the filters range from $\sim$30~mm at low temperature to $\sim$130~mm for the warmer and outer THF \cite{Barbera2018a}.

The WFI array of DEPFET active pixels are also sensitive to UV/VIS photons with energy larger that the Si band gap ($\approx$1.1~eV). The detection of UV/VIS photons degrades the detector spectral resolution, and changes the energy scale by about 3.7~eV, on average, for each optically generated electron-hole pair. Ideally, the WFI should not detect any optically generated electron-hole pair during a read-out integration time. An optical blocking filter (OBF) is thus needed in front of the detectors to observe X-ray sources with bright UV/VIS counterparts.
The currently investigated design of the WFI large area detector includes an on-chip OBF (90~nm of aluminium + 30~nm of \chem{Si_3N_4} + 20~nm of \chem{SiO_2}) and an additional OBF mounted on a filter wheel consisting of a polyimide film 150~nm thick coated with 30~nm of aluminium. The FW OBF, located nearly 100~mm far from the focal plane will be nearly 160~mm~x~160 mm in size and will be supported by a honeycomb metal mesh with $\sim$~5~mm pitch, and \SI{75}{\micro\metre} x \SI{150}{\micro\metre} bars (width x thickness) with ~3.5\% blocking factor. A cross shaped aluminum structure, overlapping the gap area between the four quadrants of the large detector array, is presently foreseen to provide additional stiffening to the very large area LDA OBF \cite{Barbera2018b}.

The current Lynx design, one of the four strategic mission concepts under study for the NASA 2020 Astrophysics Decadal Survey, includes a large telescope with 3~m diameter and 10~m focal length providing subarcsecond angular resolution coupled to three selectable focal plane instruments \cite{Gaskin2019}.  Two of the detectors, based on silicon technology, will be dedicated to high-resolution imaging and read-out of grating dispersed spectra. The third focal plane detector will be an array of TES microcalorimeters \cite{Bandler2019}. Both the semiconductor detectors will be equipped with an optical blocking filter, while for the  microcalorimeter array the use of a set of five or six thermal filters is expected. Three or four of them will split a total of 40~nm Al/80~nm polyimide supported by Si meshes, and two of them will be film-less metal fine meshes working as waveguide cutoff filters. This proposed baseline design is considered very challenging and the capability to meet the out-of-band blocking  requirements with such thin Al layers as well as the capability to manufacture 20~nm thick large area polyimide films need to be demonstrated during the Lynx development program \cite{Eckart2019}.  

Beside pushing the present technology to the limit, a new interesting technology based on CNT pellicles and carbon yarn meshes is being investigated by our group (Barbera et al. 2022 in preparation), and will be challenged in the phase B of Athena. Preliminary results show that very thin and robust CNT pellicles can be manufactured with large area and properly bonded to fine meshes. The capability to obtain a compact, stable, and highly reflective Al coating capable to meet the required  out-of-band attenuation is presently under investigation.

\acknowledgement{The authors acknowledge all the members of the research group at INAF-OAPA and UNIPA with whom they collaborate on the design and development of filters for high-energy astrophysics Space missions: Roberto Candia, Alfonso Collura, Fabio D'Anca, Gaspare Di Cicca, Nicola Montinaro, Elena Puccio, Michela Todaro, Salvatore Varisco. MB takes this opportunity to thank the late Martin Zombeck, a brilliant scientist, a close friend and a mentor, who has introduced him to this interesting subject during the development of the Chandra X-ray observatory.}     

\bibliographystyle{aa_mod}
\bibliography{biblio.bib}

\begin{thebibliography}{149}
\expandafter\ifx\csname natexlab\endcsname\relax\def\natexlab#1{#1}\fi

\bibitem[{Ando {et~al.}(2002)Ando, Watanabe, \& Matsuura}]{Ando2002}
Ando, S., Watanabe, Y., \& Matsuura, T. 2002, Jpn. J. Appl. Phys. (2008), 41,
  5254

\bibitem[{Aucoin {et~al.}(1994)Aucoin, Markert, Nenonen, Flanagan, \&
  Barbera}]{Aucoin1994}
Aucoin, R.~J., Markert, T.~H., Nenonen, S. A.~A., Flanagan, K.~A., \& Barbera,
  M. 1994, in {EUV}, X-Ray, and Gamma-Ray Instrumentation for Astronomy V, ed.
  O.~H.~W. Siegmund \& J.~V. Vallerga ({SPIE})

\bibitem[{Bandler {et~al.}(2019)Bandler, Chervenak, Datesman, Devasia, DiPirro,
  Sakai, Smith, Stevenson, Yoon, Bennett, Mates, Swetz, Ullom, Irwin, \&
  Eckart}]{Bandler2019}
Bandler, S.~R., Chervenak, J.~A., Datesman, A.~M., {et~al.} 2019, Journal of
  Astronomical Telescopes, Instruments, and Systems, 5, 1

\bibitem[{Banks {et~al.}(2004)Banks, de~Groh, \& Miller}]{Banks2004}
Banks, B.~A., de~Groh, K.~K., \& Miller, S.~K. 2004, {MRS} Proceedings, 851

\bibitem[{Barbera {et~al.}(2016)Barbera, Argan, Bozzo, Branduardi-Raymont,
  Ciaravella, Collura, Cuttaia, Gatti, Escobar, Cicero, Lotti, Macculi, Mineo,
  Nuzzo, Paltani, Parodi, Piro, Rauw, Sciortino, Sciortino, \&
  Villa}]{Barbera2016}
Barbera, M., Argan, A., Bozzo, E., {et~al.} 2016, Journal of Low Temperature
  Physics, 184, 706

\bibitem[{{Barbera} {et~al.}(1994){Barbera}, {Austin}, {Collura}, {Flanagan},
  {Jelinsky}, {Murray}, {Serio}, \& {Zombeck}}]{Barbera1994}
{Barbera}, M., {Austin}, G.~K., {Collura}, A., {et~al.} 1994, in Society of
  Photo-Optical Instrumentation Engineers (SPIE) Conference Series, Vol. 2280,
  EUV, X-Ray, and Gamma-Ray Instrumentation for Astronomy V, ed. O.~H.
  {Siegmund} \& J.~V. {Vallerga}, 214--228

\bibitem[{Barbera {et~al.}(2006)Barbera, Candia, Collura, Cicca, Pelliciari,
  Sciortino, \& Varisco}]{Barbera2006}
Barbera, M., Candia, R., Collura, A., {et~al.} 2006, in Space Telescopes and
  Instrumentation {II}: Ultraviolet to Gamma Ray, ed. M.~J.~L. Turner \&
  G.~Hasinger ({SPIE})

\bibitem[{Barbera {et~al.}(1997)Barbera, Collura, Dara, Leone, Powell, Serio,
  Varisco, \& Zombeck}]{Barbera1997}
Barbera, M., Collura, A., Dara, A., {et~al.} 1997, Experimental Astronomy, 7,
  51

\bibitem[{{Barbera} {et~al.}(2018{\natexlab{a}}){Barbera}, {Lo Cicero},
  {Sciortino}, {D'Anca}, {Lo Cicero}, {Parodi}, {Sciortino}, {Rauw},
  {Branduardi-Raymont}, {Varisco}, {Ferruggia Bonura}, {Collura}, {Candia}, {Di
  Cicca}, {Giglio}, {Buttacavoli}, {Cuttaia}, {Villa}, {Cappi}, {Lam Trong},
  {Mesnager}, {Peille}, {den Hartog}, {den Herder}, {Jackson}, {Barret}, \&
  {Piro}}]{Barbera2018a}
{Barbera}, M., {Lo Cicero}, U., {Sciortino}, L., {et~al.} 2018{\natexlab{a}},
  in Society of Photo-Optical Instrumentation Engineers (SPIE) Conference
  Series, Vol. 10699, Space Telescopes and Instrumentation 2018: Ultraviolet to
  Gamma Ray, ed. J.-W.~A. {den Herder}, S.~{Nikzad}, \& K.~{Nakazawa}, 106991R

\bibitem[{{Barbera} {et~al.}(2018{\natexlab{b}}){Barbera}, {Lo Cicero},
  {Sciortino}, {D'Anca}, {Parodi}, {Rataj}, {Polak}, {Pilch}, {Meidinger},
  {Sciortino}, {Rauw}, {Branduardi Raymont}, {Mineo}, {Perinati}, {Giglio},
  {Collura}, {Varisco}, \& {Candia}}]{Barbera2018b}
{Barbera}, M., {Lo Cicero}, U., {Sciortino}, L., {et~al.} 2018{\natexlab{b}},
  in Society of Photo-Optical Instrumentation Engineers (SPIE) Conference
  Series, Vol. 10699, Space Telescopes and Instrumentation 2018: Ultraviolet to
  Gamma Ray, ed. J.-W.~A. {den Herder}, S.~{Nikzad}, \& K.~{Nakazawa}, 106991K

\bibitem[{Barbera {et~al.}(2000)Barbera, Micela, Collura, Murray, \&
  Zombeck}]{Barbera2000}
Barbera, M., Micela, G., Collura, A., Murray, S.~S., \& Zombeck, M.~V. 2000,
  The Astrophysical Journal, 545, 449

\bibitem[{Barcons {et~al.}(2015)Barcons, Nandra, Barret, den Herder, Fabian,
  Piro, \& and}]{Barcons2015}
Barcons, X., Nandra, K., Barret, D., {et~al.} 2015, Journal of Physics:
  Conference Series, 610, 012008

\bibitem[{{Barret} {et~al.}(2018){Barret}, {Lam Trong}, {den Herder}, {Piro},
  {Cappi}, {Houvelin}, {Kelley}, {Mas-Hesse}, {Mitsuda}, {Paltani}, {Rauw},
  {Rozanska}, {Wilms}, {Bandler}, {Barbera}, {Barcons}, {Bozzo}, {Ceballos},
  {Charles}, {Costantini}, {Decourchelle}, {den Hartog}, {Duband}, {Duval},
  {Fiore}, {Gatti}, {Goldwurm}, {Jackson}, {Jonker}, {Kilbourne}, {Macculi},
  {Mendez}, {Molendi}, {Orleanski}, {Pajot}, {Pointecouteau}, {Porter},
  {Pratt}, {Pr{\^e}le}, {Ravera}, {Sato}, {Schaye}, {Shinozaki}, {Thibert},
  {Valenziano}, {Valette}, {Vink}, {Webb}, {Wise}, {Yamasaki}, {Douchin},
  {Mesnager}, {Pontet}, {Pradines}, {Branduardi-Raymont}, {Bulbul}, {Dadina},
  {Ettori}, {Finoguenov}, {Fukazawa}, {Janiuk}, {Kaastra}, {Mazzotta},
  {Miller}, {Miniutti}, {Naze}, {Nicastro}, {Scioritino}, {Simonescu},
  {Torrejon}, {Frezouls}, {Geoffray}, {Peille}, {Aicardi}, {Andr{\'e}},
  {Daniel}, {Cl{\'e}net}, {Etcheverry}, {Gloaguen}, {Hervet}, {Jolly}, {Ledot},
  {Paillet}, {Schmisser}, {Vella}, {Damery}, {Boyce}, {Dipirro}, {Lotti},
  {Schwander}, {Smith}, {Van Leeuwen}, {van Weers}, {Clerc}, {Cobo}, {Dauser},
  {Kirsch}, {Cucchetti}, {Eckart}, {Ferrando}, \& {Natalucci}}]{Barret2018}
{Barret}, D., {Lam Trong}, T., {den Herder}, J.-W., {et~al.} 2018, in Society
  of Photo-Optical Instrumentation Engineers (SPIE) Conference Series, Vol.
  10699, Space Telescopes and Instrumentation 2018: Ultraviolet to Gamma Ray,
  ed. J.-W.~A. {den Herder}, S.~{Nikzad}, \& K.~{Nakazawa}, 106991G

\bibitem[{Bautz {et~al.}(2016)Bautz, Kissel, Masterson, Ryu, \&
  Suntharalingam}]{Bautz2016}
Bautz, M., Kissel, S., Masterson, R., Ryu, K., \& Suntharalingam, V. 2016, in
  Space Telescopes and Instrumentation 2016: Ultraviolet to Gamma Ray, ed.
  J.-W.~A. den Herder, T.~Takahashi, \& M.~Bautz ({SPIE})

\bibitem[{{Bavdaz} {et~al.}(2010){Bavdaz}, {Collon}, {Beijersbergen},
  {Wallace}, \& {Wille}}]{Bavdaz2010}
{Bavdaz}, M., {Collon}, M., {Beijersbergen}, M., {Wallace}, K., \& {Wille}, E.
  2010, X-Ray Optics and Instrumentation, 2010, 295095

\bibitem[{Bhandarkar {et~al.}(2016)Bhandarkar, Betcher, Smith, Lairson, \&
  Ayers}]{Bhandarkar2016}
Bhandarkar, S., Betcher, J., Smith, R., Lairson, B., \& Ayers, T. 2016, Fusion
  Science and Technology, 70, 332

\bibitem[{Briel \& Pfeffermann(1995)}]{Briel1995}
Briel, U.~G. \& Pfeffermann, E. 1995, in {EUV}, X-Ray, and Gamma-Ray
  Instrumentation for Astronomy {VI}, ed. O.~H.~W. Siegmund \& J.~V. Vallerga
  ({SPIE})

\bibitem[{Burrows {et~al.}(2004)Burrows, Hill, Nousek, Wells, Chincarini,
  Abbey, Beardmore, Bosworth, Brauninger, Burkert, Campana, Capalbi, Chang,
  Citterio, Freyberg, Giommi, Hartner, Killough, Kittle, Klar, Mangels,
  McMeekin, Miles, Moretti, Mori, Morris, Mukerjee, Osborne, Short,
  Tagliaferri, Tamburelli, Watson, Willingale, \& Zugger}]{Burrows2004}
Burrows, D.~N., Hill, J.~E., Nousek, J.~A., {et~al.} 2004, in X-Ray and
  Gamma-Ray Instrumentation for Astronomy {XIII} ({SPIE})

\bibitem[{Carpenter {et~al.}(2006)Carpenter, Abbey, Ambrosi, \&
  Wells}]{Carpenter2006}
Carpenter, J.~D., Abbey, A.~F., Ambrosi, R.~M., \& Wells, A. 2006, in Space
  Telescopes and Instrumentation {II}: Ultraviolet to Gamma Ray, ed. M.~J.~L.
  Turner \& G.~Hasinger ({SPIE})

\bibitem[{Castelli {et~al.}(1997)Castelli, Watson, Wells, Kent, Barbera,
  Collura, \& Bavdaz}]{Castelli1997}
Castelli, C.~M., Watson, D.~J., Wells, A.~A., {et~al.} 1997, in {EUV}, X-Ray,
  and Gamma-Ray Instrumentation for Astronomy {VIII}, ed. O.~H.~W. Siegmund \&
  M.~A. Gummin ({SPIE})

\bibitem[{Chartas {et~al.}(1996)Chartas, Garmire, Nousek, Townsley, Powell,
  Blake, \& Graessle}]{Chartas1996}
Chartas, G., Garmire, G.~P., Nousek, J.~A., {et~al.} 1996, in Multilayer and
  Grazing Incidence X-Ray/{EUV} Optics {III}, ed. R.~B. Hoover \& A.~B. C.~W.
  II ({SPIE})

\bibitem[{{Collon} {et~al.}(2019){Collon}, {Vacanti}, {Barri{\`e}re},
  {Landgraf}, {G{\"u}nther}, {Vervest}, {Voruz}, {Verhoeckx}, {Babi{\'c}},
  {Keek}, {Girou}, {Okma}, {Hauser}, {Beijersbergen}, {Bavdaz}, {Wille},
  {Fransen}, {Shortt}, {Ferreira}, {Haneveld}, {Koelewijn}, {Start},
  {Wijnperle}, {Lankwarden}, {van Baren}, {Hieltjes}, {den Herder},
  {M{\"u}ller}, {Handick}, {Krumrey}, {Bradshaw}, {Burwitz}, {Pareschi},
  {Massahi}, {Svendsen}, {Della Monica Ferreira}, {Christensen}, {Valsecchi},
  {Oliver}, {Chequer}, \& {Ball}}]{Collon2019}
{Collon}, M.~J., {Vacanti}, G., {Barri{\`e}re}, N.~M., {et~al.} 2019, in
  Society of Photo-Optical Instrumentation Engineers (SPIE) Conference Series,
  Vol. 11119, Optics for EUV, X-Ray, and Gamma-Ray Astronomy IX, 111190L

\bibitem[{Collura {et~al.}(2009)Collura, Barbera, Varisco, Basso, Pareschi,
  Tagliaferri, Ayers, Rodriguez, \& Ferrando}]{Collura2009}
Collura, A., Barbera, M., Varisco, S., {et~al.} 2009, in {AIP} Conference
  Proceedings ({AIP})

\bibitem[{Cooper {et~al.}(2008)Cooper, Upadhyaya, Minton, Berman, Du, \&
  George}]{Cooper2008}
Cooper, R., Upadhyaya, H.~P., Minton, T.~K., {et~al.} 2008, Thin Solid Films,
  516, 4036

\bibitem[{{Cui} {et~al.}(2020){Cui}, {Bregman}, {Bruijn}, {Chen}, {Chen},
  {Cui}, {Fang}, {Gao}, {Gao}, {Gao}, {Gottardi}, {Gu}, {Guo}, {Guo}, {He},
  {He}, {den Herder}, {Huang}, {Li}, {Li}, {Li}, {Li}, {Li}, {Li}, {Liang},
  {Liang}, {Liang}, {Liu}, {Liu}, {Liu}, {Jaeckel}, {Ji}, {Ji}, {Jin}, {Kang},
  {Ma}, {McCammon}, {Mo}, {Nagayoshi}, {Nelms}, {Qi}, {Quan}, {Ridder}, {Shen},
  {Simionescu}, {Taralli}, {Wang}, {Wang}, {Wang}, {Wang}, {Wang}, {Wang},
  {Wang}, {Wang}, {Wang}, {Wang}, {Wang}, {Wang}, {Wang}, {Wang}, {Wen}, {de
  Wit}, {Wu}, {Xu}, {Xu}, {Xu}, {Xu}, {Xu}, {Xue}, {Yi}, {Yu}, {Yang}, {Yuan},
  {Zhang}, {Zhang}, {Zhang}, {Zhong}, {Zhou}, \& {Zhu}}]{Cui2020}
{Cui}, W., {Bregman}, J.~N., {Bruijn}, M.~P., {et~al.} 2020, in Society of
  Photo-Optical Instrumentation Engineers (SPIE) Conference Series, Vol. 11444,
  Society of Photo-Optical Instrumentation Engineers (SPIE) Conference Series,
  114442S

\bibitem[{{de Korte} {et~al.}(1981){de Korte}, Bleeker, {den Boggende},
  Branduardi-Raymont, Brinkman, Culhane, Gronenschild, Mason, \&
  McKechnie}]{deKorte1981}
{de Korte}, P. A.~J., Bleeker, J. A.~M., {den Boggende}, A. J.~F., {et~al.}
  1981, Space Science Reviews, 30, 495

\bibitem[{de~Vries {et~al.}(2017)de~Vries, Haas, Yamasaki, den Herder, Paltani,
  Kilbourne, Tsujimoto, Eckart, \& Leutenegger}]{DeVries2017}
de~Vries, C.~P., Haas, D., Yamasaki, N.~Y., {et~al.} 2017, Journal of
  Astronomical Telescopes, Instruments, and Systems, 4, 1

\bibitem[{den Herder {et~al.}(2001)den Herder, Brinkman, Kahn,
  Branduardi-Raymont, Thomsen, Aarts, Audard, Bixler, den Boggende, Cottam,
  Decker, Dubbeldam, Erd, Goulooze, G\"{u}del, Guttridge, Hailey, Janabi,
  Kaastra, de~Korte, van Leeuwen, Mauche, McCalden, Mewe, Naber, Paerels,
  Peterson, Rasmussen, Rees, Sakelliou, Sako, Spodek, Stern, Tamura, Tandy,
  de~Vries, Welch, \& Zehnder}]{denherder2001}
den Herder, J.~W., Brinkman, A.~C., Kahn, S.~M., {et~al.} 2001, Astronomy {\&}
  Astrophysics, 365, L7

\bibitem[{DeOliveira {et~al.}(2012)DeOliveira, Albuquerque, Cruz, Yamaji, \&
  Leite}]{Deoliveira2012}
DeOliveira, R., Albuquerque, D., Cruz, T., Yamaji, F., \& Leite, F. 2012, in
  Atomic Force Microscopy - Imaging, Measuring and Manipulating Surfaces at the
  Atomic Scale ({InTech}), 147--174

\bibitem[{Eckart {et~al.}(2018)Eckart, Adams, Boyce, Brown, Chiao, Fujimoto,
  Haas, den Herder, Hoshino, Ishisaki, Kilbourne, Kitamoto, Leutenegger,
  McCammon, Mitsuda, Porter, Sato, Sawada, Seta, Sneiderman, Szymkowiak, Takei,
  Tashiro, Tsujimoto, de~Vries, Watanabe, Yamada, \& Yamasaki}]{Eckart2018}
Eckart, M.~E., Adams, J.~S., Boyce, K.~R., {et~al.} 2018, Journal of
  Astronomical Telescopes, Instruments, and Systems, 4, 1

\bibitem[{Eckart \& Yoon(2019)}]{Eckart2019}
Eckart, M.~E. \& Yoon, W. 2019, Journal of Astronomical Telescopes,
  Instruments, and Systems, 5, 1

\bibitem[{Forouhi \& Bloomer(1986)}]{Forouhi1986}
Forouhi, A.~R. \& Bloomer, I. 1986, Phys. Rev. B, 34, 7018

\bibitem[{Forouhi \& Bloomer(2019)}]{Forouhi2019}
Forouhi, A.~R. \& Bloomer, I. 2019, Journal of Physics Communications, 3,
  035022

\bibitem[{Frey \& Khan(2015)}]{Frey2015}
Frey, H. \& Khan, H.~R. 2015, Handbook of thin film technology (Springer)

\bibitem[{Freyberg {et~al.}(2006)Freyberg, Br\"{a}uninger, Burkert, Hartner,
  Citterio, Mazzoleni, Pareschi, Spiga, Romaine, Gorenstein, \&
  Ramsey}]{Freyberg2006}
Freyberg, M.~J., Br\"{a}uninger, H., Burkert, W., {et~al.} 2006, Experimental
  Astronomy, 20, 405

\bibitem[{Friedrich {et~al.}(1998)Friedrich, Braeuninger, Burkert, Doehring,
  Egger, Hasinger, Oppitz, Predehl, \& Truemper}]{Friedrich1998}
Friedrich, P., Braeuninger, H.~W., Burkert, W., {et~al.} 1998, in X-Ray Optics,
  Instruments, and Missions, ed. R.~B. Hoover \& A.~B. C.~W. II ({SPIE})

\bibitem[{Garmire {et~al.}(2003)Garmire, Bautz, Ford, Nousek, \& George
  R.~Ricker}]{Garmire2003}
Garmire, G.~P., Bautz, M.~W., Ford, P.~G., Nousek, J.~A., \& George R.~Ricker,
  J. 2003, in X-Ray and Gamma-Ray Telescopes and Instruments for Astronomy, ed.
  J.~E. Truemper \& H.~D. Tananbaum ({SPIE})

\bibitem[{Gaskin \& Swartz(2019)}]{Gaskin2019}
Gaskin, J.~A. \& Swartz, D.~A. 2019, Journal of Astronomical Telescopes,
  Instruments, and Systems, 5, 1

\bibitem[{{Gehrels} {et~al.}(2004){Gehrels}, {Chincarini}, {Giommi}, {Mason},
  {Nousek}, {Wells}, {White}, {Barthelmy}, {Burrows}, {Cominsky}, {Hurley},
  {Marshall}, {M{\'e}sz{\'a}ros}, {Roming}, {Angelini}, {Barbier}, {Belloni},
  {Campana}, {Caraveo}, {Chester}, {Citterio}, {Cline}, {Cropper}, {Cummings},
  {Dean}, {Feigelson}, {Fenimore}, {Frail}, {Fruchter}, {Garmire}, {Gendreau},
  {Ghisellini}, {Greiner}, {Hill}, {Hunsberger}, {Krimm}, {Kulkarni}, {Kumar},
  {Lebrun}, {Lloyd-Ronning}, {Markwardt}, {Mattson}, {Mushotzky}, {Norris},
  {Osborne}, {Paczynski}, {Palmer}, {Park}, {Parsons}, {Paul}, {Rees},
  {Reynolds}, {Rhoads}, {Sasseen}, {Schaefer}, {Short}, {Smale}, {Smith},
  {Stella}, {Tagliaferri}, {Takahashi}, {Tashiro}, {Townsley}, {Tueller},
  {Turner}, {Vietri}, {Voges}, {Ward}, {Willingale}, {Zerbi}, \&
  {Zhang}}]{Gehrels2004}
{Gehrels}, N., {Chincarini}, G., {Giommi}, P., {et~al.} 2004, Astrophysical
  Journal, 611, 1005

\bibitem[{Giacconi {et~al.}(1979)Giacconi, Branduardi, Briel, Epstein,
  Fabricant, Feigelson, Forman, Gorenstein, Grindlay, Gursky,
  {et~al.}}]{Giacconi1979}
Giacconi, R., Branduardi, G., Briel, U., {et~al.} 1979, The Astrophysical
  Journal, 230, 540

\bibitem[{Golub(1990)}]{Spiller1990}
Golub, L. 1990, Optical Engineering, 29, 625

\bibitem[{Golub {et~al.}(2007)Golub, DeLuca, Austin, Bookbinder, Caldwell,
  Cheimets, Cirtain, Cosmo, Reid, Sette, Weber, Sakao, Kano, Shibasaki, Hara,
  Tsuneta, Kumagai, Tamura, Shimojo, McCracken, Carpenter, Haight, Siler,
  Wright, Tucker, Rutledge, Barbera, Peres, \& Varisco}]{Golub2007}
Golub, L., DeLuca, E., Austin, G., {et~al.} 2007, Solar Physics, 243, 63

\bibitem[{Gorenstein {et~al.}(1981)Gorenstein, Harnden, \&
  Fabricant}]{Gorenstein1981}
Gorenstein, P., Harnden, F.~R., \& Fabricant, D.~G. 1981, {IEEE} Transactions
  on Nuclear Science, 28, 869

\bibitem[{Gottwald {et~al.}(2006)Gottwald, Kroth, Krumrey, Richter, Scholze, \&
  Ulm}]{Gottwald2006}
Gottwald, A., Kroth, U., Krumrey, M., {et~al.} 2006, Metrologia, 43, S125

\bibitem[{Grove(1999)}]{Grove1999}
Grove, D.~A. 1999, in X-Ray Optics, Instruments, and Missions {II}, ed. R.~B.
  Hoover \& A.~B. C.~W. II ({SPIE})

\bibitem[{Grove {et~al.}(2010)Grove, Betcher, Lairson, Smith, \&
  Ayers}]{Grove2010}
Grove, D.~A., Betcher, J.~C., Lairson, B., Smith, R., \& Ayers, T. 2010, in
  Modern Technologies in Space- and Ground-based Telescopes and Instrumentation
  ({SPIE})

\bibitem[{Hagemann {et~al.}(1975)Hagemann, Gudat, \& Kunz}]{Hagemann1975}
Hagemann, H.-J., Gudat, W., \& Kunz, C. 1975, Journal of the Optical Society of
  America, 65, 742

\bibitem[{Hasegawa \& Horie(2001)}]{Hasegawa2001}
Hasegawa, M. \& Horie, K. 2001, Progress in Polymer Science, 26, 259

\bibitem[{Heavens(1960)}]{Heavens1960}
Heavens, O.~S. 1960, Reports on Progress in Physics, 23, 1

\bibitem[{Heitler(1954)}]{Heitler1954}
Heitler, W. 1954, The quantum theory of radiation (Oxford: Clarendon Press),
  204--211

\bibitem[{Henke {et~al.}(1993)Henke, Gullikson, \& Davis}]{Henke1993}
Henke, B., Gullikson, E., \& Davis, J. 1993, Atomic Data and Nuclear Data
  Tables, 54, 181

\bibitem[{Henry {et~al.}(1977)Henry, Kellogg, Briel, Murray, Speybroeck, \&
  Bjorkholm}]{Henry1977}
Henry, J.~P., Kellogg, E.~M., Briel, U.~G., {et~al.} 1977, in X-Ray Imaging,
  ed. R.~P. Chase \& G.~W. Kuswa ({SPIE})

\bibitem[{Hill(1994)}]{Hill1994}
Hill, D. 1994, {IEEE} Transactions on Electromagnetic Compatibility, 36, 294

\bibitem[{Hubbell(1969)}]{Hubbell1969}
Hubbell, J.~H. 1969, National Bureau of Standards Report NSRDS-NBS29,
  Washington DC

\bibitem[{Hubbell(2006)}]{Hubbell2006}
Hubbell, J.~H. 2006, Physics in Medicine and Biology, 51, R245

\bibitem[{Hubbell {et~al.}(1975)Hubbell, Veigele, Briggs, Brown, Cromer, \&
  Howerton}]{Hubbell1975}
Hubbell, J.~H., Veigele, W.~J., Briggs, E.~A., {et~al.} 1975, Journal of
  Physical and Chemical Reference Data, 4, 471

\bibitem[{Huebner {et~al.}(2016)Huebner, Miyakawa, Pahlke, \&
  Kreupl}]{Huebner2016}
Huebner, S., Miyakawa, N., Pahlke, A., \& Kreupl, F. 2016, {MRS} Advances, 1,
  1441

\bibitem[{Idir {et~al.}(2010)Idir, Mercere, Moreno, Delmotte, Dasilva, Modi,
  Garrett, Gentle, Nugent, \& Wilkins}]{Idir2010}
Idir, M., Mercere, P., Moreno, T., {et~al.} 2010, in AIP Conference Proceedings
  ({AIP})

\bibitem[{Jablonski(2019)}]{Jablonski2019}
Jablonski, A. 2019, Surface Science, 688, 14

\bibitem[{Kawka \& Buckius(2001)}]{Kawka2001}
Kawka, P.~A. \& Buckius, R.~O. 2001, International Journal of Thermophysics,
  22, 517

\bibitem[{Kelley {et~al.}(2016)Kelley, Akamatsu, Azzarello, Bialas, Boyce,
  Brown, Canavan, Chiao, Costantini, DiPirro, Eckart, Ezoe, Fujimoto, Haas, den
  Herder, Hoshino, Ishikawa, Ishisaki, Iyomoto, Kilbourne, Kimball, Kitamoto,
  Konami, Koyama, Leutenegger, McCammon, Mitsuda, Mitsuishi, Moseley, Murakami,
  Murakami, Noda, Ogawa, Ohashi, Okamoto, Ota, Paltani, Porter, Sakai, Sato,
  Sato, Sawada, Seta, Shinozaki, Shirron, Sneiderman, Sugita, Szymkowiak,
  Takei, Tamagawa, Tashiro, Terada, Tsujimoto, de~Vries, Yamada, Yamasaki, \&
  Yatsu}]{Kelley2016}
Kelley, R.~L., Akamatsu, H., Azzarello, P., {et~al.} 2016, in Space Telescopes
  and Instrumentation 2016: Ultraviolet to Gamma Ray, ed. J.-W.~A. den Herder,
  T.~Takahashi, \& M.~Bautz ({SPIE})

\bibitem[{Kelley {et~al.}(2007)Kelley, Mitsuda, Allen, Arsenovic, Audley,
  Bialas, Boyce, Boyle, Breon, Brown, Cottam, DiPirro, Fujimoto, Furusho,
  Gendreau, Gochar, Gonzalez, Hirabayashi, Holt, Inoue, Ishida, Ishisaki,
  Jones, Keski-Kuha, Kilbourne, McCammon, Morita, Moseley, Mott, Narasaki,
  Ogawara, Ohashi, Ota, Panek, Porter, Serlemitsos, Shirron, Sneiderman,
  Szymkowiak, Takei, Tveekrem, Volz, Yamamoto, \& Yamasaki}]{Kelley2007}
Kelley, R.~L., Mitsuda, K., Allen, C.~A., {et~al.} 2007, Publications of the
  Astronomical Society of Japan, 59, S77

\bibitem[{{Kenter} {et~al.}(2000){Kenter}, {Chappell}, {Kraft}, {Meehan},
  {Murray}, {Zombeck}, {Hole}, {Juda}, {Donnelly}, {Patnaude}, {Pease},
  {Wilton}, {Zhao}, {Austin}, {Fraser}, {Pearson}, {Lees}, {Brunton},
  {Barbera}, {Collura}, \& {Serio}}]{Kenter2000}
{Kenter}, A.~T., {Chappell}, J.~H., {Kraft}, R.~P., {et~al.} 2000, in Society
  of Photo-Optical Instrumentation Engineers (SPIE) Conference Series, Vol.
  4012, X-Ray Optics, Instruments, and Missions III, ed. J.~E. {Truemper} \&
  B.~{Aschenbach}, 467--492

\bibitem[{Kilbourne {et~al.}(2018)Kilbourne, Adams, Arsenovic, Ayers, Chiao,
  DiPirro, Eckart, Fujimoto, Kazeva, Kripps, Lairson, Leutenegger, Lopez,
  McCammon, McGuinness, Mitsuda, Moseley, Porter, Schweiss, \&
  Takei}]{Kilbourne2018}
Kilbourne, C.~A., Adams, J.~S., Arsenovic, P., {et~al.} 2018, Journal of
  Astronomical Telescopes, Instruments, and Systems, 4, 1

\bibitem[{Kim {et~al.}(2008)Kim, Kim, Huh, Yang, Kim, Ha, Won, Kim, \&
  Lee}]{Kim2008}
Kim, D.-W., Kim, D.-I., Huh, Y.-H., {et~al.} 2008, Nuclear Instruments and
  Methods in Physics Research Section B: Beam Interactions with Materials and
  Atoms, 266, 3263

\bibitem[{Kirsch {et~al.}(2005)Kirsch, Abbey, Altieri, Baskill, Dennerl, van
  Dooren, Fauste, Freyberg, Gabriel, Haberl, Hartmann, Hartner, Meidinger,
  Metcalfe, Olabarri, Pollock, Read, Rives, Sembay, Smith, Stuhlinger, \&
  Talavera}]{Kirsch2005}
Kirsch, M.~G., Abbey, A., Altieri, B., {et~al.} 2005, in {UV}, X-Ray, and
  Gamma-Ray Space Instrumentation for Astronomy {XIV}, ed. O.~H.~W. Siegmund
  ({SPIE})

\bibitem[{{KLA Corporation}(Accessed on: 10/2021)}]{filmetrics.com}
{KLA Corporation}. Accessed on: 10/2021, Refractive index database

\bibitem[{Kohmura {et~al.}(2000)Kohmura, Katayama, Asakura, Kitamoto, Tsunemi,
  Hayashida, Miyata, Hashimotodani, Katayama, Shouho, Koyama, Kissel, Ricker,
  Bautz, \& Foster}]{Kohmura2000}
Kohmura, T., Katayama, K., Asakura, R., {et~al.} 2000, Advances in Space
  Research, 25, 877

\bibitem[{Koningsberger \& Prins(1987)}]{Koningsberger1987}
Koningsberger, D.~C. \& Prins, R. 1987, X-ray absorption: Principles,
  applications, techniques of EXAFS, SEXAFS and XANES (John Wiley and Sons
  Inc., New York, NY)

\bibitem[{Koyama {et~al.}(2007)Koyama, Tsunemi, Dotani, Bautz, Hayashida,
  Tsuru, Matsumoto, Ogawara, Ricker, Doty, Kissel, Foster, Nakajima, Yamaguchi,
  Mori, Sakano, Hamaguchi, Nishiuchi, Miyata, Torii, Namiki, Katsuda, Matsuura,
  Miyauchi, Anabuki, Tawa, Ozaki, Murakami, Maeda, Ichikawa, Prigozhin,
  Boughan, LaMarr, Miller, Burke, Gregory, Pillsbury, Bamba, Hiraga, Senda,
  Katayama, Kitamoto, Tsujimoto, Kohmura, Tsuboi, \& Awaki}]{Koyama2007}
Koyama, K., Tsunemi, H., Dotani, T., {et~al.} 2007, Publications of the
  Astronomical Society of Japan, 59, S23

\bibitem[{Lee {et~al.}(2003)Lee, Seo, Shul, \& Han}]{Lee2003}
Lee, C., Seo, J., Shul, Y., \& Han, H. 2003, Polymer Journal, 35, 578

\bibitem[{Lee {et~al.}(2008)Lee, Wei, Kysar, \& Hone}]{Lee2008}
Lee, C., Wei, X., Kysar, J.~W., \& Hone, J. 2008, Science, 321, 385

\bibitem[{L{\'{e}}vesque {et~al.}(1994)L{\'{e}}vesque, Paton, \&
  Payne}]{Lvesque1994}
L{\'{e}}vesque, L., Paton, B.~E., \& Payne, S.~H. 1994, Applied Optics, 33,
  8036

\bibitem[{Li {et~al.}(2007)Li, Li, He, Di, \& Yang}]{Li2007}
Li, R., Li, C., He, S., Di, M., \& Yang, D. 2007, Radiation Physics and
  Chemistry, 76, 1200

\bibitem[{Li {et~al.}(2011)Li, Zhang, \& Michaux}]{Li2011}
Li, T., Zhang, Z., \& Michaux, B. 2011, Theoretical and Applied Mechanics
  Letters, 1, 041002

\bibitem[{Lo~Cicero {et~al.}(2018)Lo~Cicero, Lo~Cicero, Puccio, Montinaro,
  Gulli, Todaro, Calandra, Torma, Cuttaia, Villa, {et~al.}}]{LoCicero2018}
Lo~Cicero, U., Lo~Cicero, G., Puccio, E., {et~al.} 2018, in Space Telescopes
  and Instrumentation 2018: Ultraviolet to Gamma Ray, ed. J.-W.~A. den Herder,
  K.~Nakazawa, \& S.~Nikzad, Vol. 10699, International Society for Optics and
  Photonics ({SPIE}), 106994R

\bibitem[{Lotti {et~al.}(2017)Lotti, Mineo, Jacquey, Molendi, D'Andrea,
  Macculi, \& Piro}]{Lotti2017}
Lotti, S., Mineo, T., Jacquey, C., {et~al.} 2017, Experimental Astronomy, 44,
  371

\bibitem[{{Malina} {et~al.}(1992){Malina}, {Bowyer}, {Abbott}, {Christian},
  {Drake}, {Dupuis}, {Finley}, {Fruscione}, {Hawkins}, {Jelinsky}, {Lieu},
  {Marshall}, {Patterer}, {Vallerga}, {Vedder}, \& {Vennes}}]{Malina1992}
{Malina}, R.~F., {Bowyer}, S., {Abbott}, M., {et~al.} 1992, in American
  Astronomical Society Meeting Abstracts, Vol. 181, American Astronomical
  Society Meeting Abstracts, 23.01

\bibitem[{Manivannan(1997)}]{Manivannan1997}
Manivannan, G. 1997, in Materials Characterization and Optical Probe
  Techniques: A Critical Review ({SPIE})

\bibitem[{Marshall {et~al.}(2004)Marshall, Tennant, Grant, Hitchcock,
  O{\textquotesingle}Dell, \& Plucinsky}]{Marshall2004}
Marshall, H.~L., Tennant, A., Grant, C.~E., {et~al.} 2004, in X-Ray and
  Gamma-Ray Instrumentation for Astronomy {XIII} ({SPIE})

\bibitem[{McCammon {et~al.}(2002)McCammon, Almy, Apodaca, Tiest, Cui, Deiker,
  Galeazzi, Juda, Lesser, Mihara, Morgenthaler, Sanders, Zhang,
  Figueroa-Feliciano, Kelley, Moseley, Mushotzky, Porter, Stahle, \&
  Szymkowiak}]{McCammon2002}
McCammon, D., Almy, R., Apodaca, E., {et~al.} 2002, The Astrophysical Journal,
  576, 188

\bibitem[{{McCammon} {et~al.}(2008){McCammon}, {Barger}, {Brandl}, {Brekosky},
  {Crowder}, {Gygax}, {Kelley}, {Kilbourne}, {Lindeman}, {Porter}, {Rocks}, \&
  {Szymkowiak}}]{McCammon2008}
{McCammon}, D., {Barger}, K., {Brandl}, D.~E., {et~al.} 2008, Journal of Low
  Temperature Physics, 151, 715

\bibitem[{Meehan {et~al.}(1997)Meehan, Murray, Zombeck, Kraft, Kobayashi,
  Chappell, Kenter, Barbera, Collura, \& Serio}]{Meehan1997}
Meehan, G.~R., Murray, S.~S., Zombeck, M.~V., {et~al.} 1997, in {EUV}, X-Ray,
  and Gamma-Ray Instrumentation for Astronomy {VIII}, ed. O.~H.~W. Siegmund \&
  M.~A. Gummin ({SPIE})

\bibitem[{Meidinger {et~al.}(2020)Meidinger, Andritschke, Dennerl, Emberger,
  Eraerds, Haelker, Hartner, Pietschner, \& Reiffers}]{Meidinger2020}
Meidinger, N., Andritschke, R., Dennerl, K., {et~al.} 2020, in Space Telescopes
  and Instrumentation 2020: Ultraviolet to Gamma Ray, ed. J.-W.~A. den Herder,
  K.~Nakazawa, \& S.~Nikzad ({SPIE})

\bibitem[{Meidinger {et~al.}(2021)Meidinger, Andritschke, Dennerl, Emberger,
  Eraerds, H\"{a}lker, Hartner, Pietschner, \& Reiffers}]{Meidinger2021}
Meidinger, N., Andritschke, R., Dennerl, K., {et~al.} 2021, Journal of
  Astronomical Telescopes, Instruments, and Systems, 7

\bibitem[{{Meidinger} {et~al.}(2017){Meidinger}, {Barbera}, {Emberger},
  {F{\"u}rmetz}, {Manhart}, {M{\"u}ller-Seidlitz}, {Nandra}, {Plattner}, {Rau},
  \& {Treberspurg}}]{Meidinger2017}
{Meidinger}, N., {Barbera}, M., {Emberger}, V., {et~al.} 2017, in Society of
  Photo-Optical Instrumentation Engineers (SPIE) Conference Series, Vol. 10397,
  Society of Photo-Optical Instrumentation Engineers (SPIE) Conference Series,
  103970V

\bibitem[{{Meidinger} {et~al.}(2014){Meidinger}, {Nandra}, {Plattner}, {Porro},
  {Rau}, {Santangelo}, {Tenzer}, \& {Wilms}}]{Meidinger2014}
{Meidinger}, N., {Nandra}, K., {Plattner}, M., {et~al.} 2014, Journal of
  Astronomical Telescopes, Instruments, and Systems, 1, 014006

\bibitem[{Mitsuda {et~al.}(2007)Mitsuda, Bautz, Inoue, Kelley, Koyama, Kunieda,
  Makishima, Ogawara, Petre, Takahashi, Tsunemi, White, Anabuki, Angelini,
  Arnaud, Awaki, Bamba, Boyce, Brown, Chan, Cottam, Dotani, Doty, Ebisawa,
  Ezoe, Fabian, Figueroa, Fujimoto, Fukazawa, Furusho, Furuzawa, Gendreau,
  Griffiths, Haba, Hamaguchi, Harrus, Hasinger, Hatsukade, Hayashida, Henry,
  Hiraga, Holt, Hornschemeier, Hughes, Hwang, Ishida, Ishisaki, Isobe, Itoh,
  Iyomoto, Kahn, Kamae, Katagiri, Kataoka, Katayama, Kawai, Kilbourne,
  Kinugasa, Kissel, Kitamoto, Kohama, Kohmura, Kokubun, Kotani, Kotoku, Kubota,
  Madejski, Maeda, Makino, Markowitz, Matsumoto, Matsumoto, Matsuoka,
  Matsushita, McCammon, Mihara, Misaki, Miyata, Mizuno, Mori, Mori, Morii,
  Moseley, Mukai, Murakami, Murakami, Mushotzky, Nagase, Namiki, Negoro,
  Nakazawa, Nousek, Okajima, yasushi ogasaka, Ohashi, Oshima, Ota, Ozaki,
  Ozawa, Parmar, Pence, Porter, Reeves, Ricker, Sakurai, Sanders, Senda,
  Serlemitsos, Shibata, Soong, Smith, Suzuki, Szymkowiak, Takahashi, Tamagawa,
  Tamura, Tamura, Tanaka, Tashiro, Tawara, Terada, Terashima, Tomida, Torii,
  Tsuboi, Tsujimoto, Tsuru, Turner, Ueda, Ueno, Ueno, Uno, Urata, Watanabe,
  Yamamoto, Yamaoka, Yamasaki, Yamashita, Yamauchi, Yamauchi, Yaqoob, Yonetoku,
  \& Yoshida}]{Mitsuda2007}
Mitsuda, K., Bautz, M., Inoue, H., {et~al.} 2007, Publications of the
  Astronomical Society of Japan, 59, S1

\bibitem[{Murray {et~al.}(2000)Murray, Austin, Chappell, Gomes, Kenter, Kraft,
  Meehan, Zombeck, Fraser, \& Serio}]{Murray2000}
Murray, S.~S., Austin, G.~K., Chappell, J.~H., {et~al.} 2000, in X-Ray Optics,
  Instruments, and Missions III, ed. J.~E. Truemper \& B.~Aschenbach, Vol.
  4012, International Society for Optics and Photonics (SPIE), 68 -- 80

\bibitem[{Murray {et~al.}(1997)Murray, Chappell, Kenter, Kobayashi, Kraft,
  Meehan, Zombeck, Fraser, Pearson, Lees, Brunton, Pearce, Barbera, Collura, \&
  Serio}]{Murray1997}
Murray, S.~S., Chappell, J.~H., Kenter, A.~T., {et~al.} 1997, in {EUV}, X-Ray,
  and Gamma-Ray Instrumentation for Astronomy {VIII}, ed. O.~H.~W. Siegmund \&
  M.~A. Gummin ({SPIE})

\bibitem[{Nair {et~al.}(2008)Nair, Blake, Grigorenko, Novoselov, Booth,
  Stauber, Peres, \& Geim}]{Nair2008}
Nair, R.~R., Blake, P., Grigorenko, A.~N., {et~al.} 2008, Science, 320, 1308

\bibitem[{{Nandra} {et~al.}(2013){Nandra}, {Barret}, {Barcons}, {Fabian}, {den
  Herder}, {Piro}, {Watson}, {Adami}, {Aird}, {Afonso}, {Alexander},
  {Argiroffi}, {Amati}, {Arnaud}, {Atteia}, {Audard}, {Badenes}, {Ballet},
  {Ballo}, {Bamba}, {Bhardwaj}, {Stefano Battistelli}, {Becker}, {De Becker},
  {Behar}, {Bianchi}, {Biffi}, {B{\^\i}rzan}, {Bocchino}, {Bogdanov}, {Boirin},
  {Boller}, {Borgani}, {Borm}, {Bouch{\'e}}, {Bourdin}, {Bower}, {Braito},
  {Branchini}, {Branduardi-Raymont}, {Bregman}, {Brenneman}, {Brightman},
  {Br{\"u}ggen}, {Buchner}, {Bulbul}, {Brusa}, {Bursa}, {Caccianiga},
  {Cackett}, {Campana}, {Cappelluti}, {Cappi}, {Carrera}, {Ceballos},
  {Christensen}, {Chu}, {Churazov}, {Clerc}, {Corbel}, {Corral}, {Comastri},
  {Costantini}, {Croston}, {Dadina}, {D'Ai}, {Decourchelle}, {Della Ceca},
  {Dennerl}, {Dolag}, {Done}, {Dovciak}, {Drake}, {Eckert}, {Edge}, {Ettori},
  {Ezoe}, {Feigelson}, {Fender}, {Feruglio}, {Finoguenov}, {Fiore}, {Galeazzi},
  {Gallagher}, {Gandhi}, {Gaspari}, {Gastaldello}, {Georgakakis},
  {Georgantopoulos}, {Gilfanov}, {Gitti}, {Gladstone}, {Goosmann}, {Gosset},
  {Grosso}, {Guedel}, {Guerrero}, {Haberl}, {Hardcastle}, {Heinz}, {Alonso
  Herrero}, {Herv{\'e}}, {Holmstrom}, {Iwasawa}, {Jonker}, {Kaastra}, {Kara},
  {Karas}, {Kastner}, {King}, {Kosenko}, {Koutroumpa}, {Kraft}, {Kreykenbohm},
  {Lallement}, {Lanzuisi}, {Lee}, {Lemoine-Goumard}, {Lobban}, {Lodato},
  {Lovisari}, {Lotti}, {McCharthy}, {McNamara}, {Maggio}, {Maiolino}, {De
  Marco}, {de Martino}, {Mateos}, {Matt}, {Maughan}, {Mazzotta}, {Mendez},
  {Merloni}, {Micela}, {Miceli}, {Mignani}, {Miller}, {Miniutti}, {Molendi},
  {Montez}, {Moretti}, {Motch}, {Naz{\'e}}, {Nevalainen}, {Nicastro}, {Nulsen},
  {Ohashi}, {O'Brien}, {Osborne}, {Oskinova}, {Pacaud}, {Paerels}, {Page},
  {Papadakis}, {Pareschi}, {Petre}, {Petrucci}, {Piconcelli}, {Pillitteri},
  {Pinto}, {de Plaa}, {Pointecouteau}, {Ponman}, {Ponti}, {Porquet}, {Pounds},
  {Pratt}, {Predehl}, {Proga}, {Psaltis}, {Rafferty}, {Ramos-Ceja}, {Ranalli},
  {Rasia}, {Rau}, {Rauw}, {Rea}, {Read}, {Reeves}, {Reiprich}, {Renaud},
  {Reynolds}, {Risaliti}, {Rodriguez}, {Rodriguez Hidalgo}, {Roncarelli},
  {Rosario}, {Rossetti}, {Rozanska}, {Rovilos}, {Salvaterra}, {Salvato}, {Di
  Salvo}, {Sanders}, {Sanz-Forcada}, {Schawinski}, {Schaye}, {Schwope},
  {Sciortino}, {Severgnini}, {Shankar}, {Sijacki}, {Sim}, {Schmid}, {Smith},
  {Steiner}, {Stelzer}, {Stewart}, {Strohmayer}, {Str{\"u}der}, {Sun}, {Takei},
  {Tatischeff}, {Tiengo}, {Tombesi}, {Trinchieri}, {Tsuru}, {Ud-Doula},
  {Ursino}, {Valencic}, {Vanzella}, {Vaughan}, {Vignali}, {Vink}, {Vito},
  {Volonteri}, {Wang}, {Webb}, {Willingale}, {Wilms}, {Wise}, {Worrall},
  {Young}, {Zampieri}, {In't Zand}, {Zane}, {Zezas}, {Zhang}, \&
  {Zhuravleva}}]{Nandra2013}
{Nandra}, K., {Barret}, D., {Barcons}, X., {et~al.} 2013, arXiv e-prints,
  arXiv:1306.2307

\bibitem[{Nannarone(2004)}]{Nannarone2004}
Nannarone, S. 2004, in {AIP} Conference Proceedings ({AIP})

\bibitem[{Nelms {et~al.}(2002)Nelms, Galeazzi, Liu, McCammon, Moeckel, Sanders,
  \& Tan}]{Nelms2002}
Nelms, K.~L., Galeazzi, M., Liu, D., {et~al.} 2002, in AIP Conference
  Proceedings (American Institute of Physics)

\bibitem[{O{\textquotesingle}Dell {et~al.}(2017)O{\textquotesingle}Dell,
  Swartz, Tice, Plucinsky, Marshall, Bogdan, Grant, Tennant, \&
  Dahmer}]{ODell2017}
O{\textquotesingle}Dell, S.~L., Swartz, D.~A., Tice, N.~W., {et~al.} 2017, in
  {UV}, X-Ray, and Gamma-Ray Space Instrumentation for Astronomy {XX}, ed.
  O.~H. Siegmund ({SPIE})

\bibitem[{O{\textquotesingle}Dell \& Weisskopf(1998)}]{ODell1998}
O{\textquotesingle}Dell, S.~L. \& Weisskopf, M.~C. 1998, in X-Ray Optics,
  Instruments, and Missions, ed. R.~B. Hoover \& A.~B. C.~W. II ({SPIE})

\bibitem[{Ohashi {et~al.}(1996)Ohashi, Ebisawa, Fukazawa, Hiyoshi, Horii,
  Ikebe, Ikeda, Inoue, Ishida, Ishisaki, Ishizuka, Kamijo, Kaneda, Kohmura,
  Makishima, Mihara, Tashiro, Murakami, Shoumura, Tanaka, Ueda, Taguchi, Tsuru,
  \& Takeshima}]{Ohashi1996}
Ohashi, T., Ebisawa, K., Fukazawa, Y., {et~al.} 1996, Publications of the
  Astronomical Society of Japan, 48, 157

\bibitem[{Paerels {et~al.}(1990)Paerels, Brinkman, den Boggende, de~Korte, \&
  Dijkstra}]{Paerels1990}
Paerels, F., Brinkman, A., den Boggende, A., de~Korte, P., \& Dijkstra, J.
  1990, Astronomy and Astrophysics Supplement Series, 85, 1021

\bibitem[{{Pajot F.}(2019)}]{Pajot2019}
{Pajot F.} 2019, {X-IFU} calibration plan, {XIFU-PL-XCAT-180626-IRAP}, Tech.
  rep., X-IFU Consortium

\bibitem[{Palik(1997)}]{Palik1997}
Palik, E.~D. 1997, Handbook of optical constants of solids (Academic press)

\bibitem[{Parmar {et~al.}(1997)Parmar, Martin, Bavdaz, Favata, Kuulkers,
  Vacanti, Lammers, Peacock, \& Taylor}]{Parmar1997}
Parmar, A.~N., Martin, D.~D., Bavdaz, M., {et~al.} 1997, Astronomy and
  Astrophysics Supplement Series, 122, 309

\bibitem[{Perinati {et~al.}(2017)Perinati, Rott, Santangelo, \&
  Tenzer}]{Perinati2017}
Perinati, E., Rott, M., Santangelo, A., \& Tenzer, C. 2017, Experimental
  Astronomy, 44, 337

\bibitem[{Pfeffermann {et~al.}(1986)Pfeffermann, Briel, Hippmann, Kettenring,
  Metzner, Predehl, Reger, Stephan, Zombeck, Chappell, \&
  Murray}]{Pfeffermann1986}
Pfeffermann, E., Briel, U.~G., Hippmann, H., {et~al.} 1986, in Soft X-Ray
  Optics and Technology, ed. E.~Koch \& G.~A. Schmahl ({SPIE})

\bibitem[{Plucinsky {et~al.}(2016)Plucinsky, Bogdan, Germain, \&
  Marshall}]{Plucinsky2016}
Plucinsky, P.~P., Bogdan, A., Germain, G., \& Marshall, H.~L. 2016, in Space
  Telescopes and Instrumentation 2016: Ultraviolet to Gamma Ray, ed. J.-W.~A.
  den Herder, T.~Takahashi, \& M.~Bautz ({SPIE})

\bibitem[{Polyanskiy(Accessed on: 10/2021)}]{refractiveindex.info}
Polyanskiy, M.~N. Accessed on: 10/2021, Refractive index database

\bibitem[{Powell {et~al.}(1997)Powell, Keski-Kuha, Zombeck, Goddard, Chartas,
  Townsley, Moebius, Davis, \& Mason}]{Powell1997}
Powell, F.~R., Keski-Kuha, R. A.~M., Zombeck, M.~V., {et~al.} 1997, in Grazing
  Incidence and Multilayer X-Ray Optical Systems, ed. R.~B. Hoover \& A.~B.
  C.~W. II ({SPIE})

\bibitem[{Predehl {et~al.}(2021)Predehl, Andritschke, Arefiev, Babyshkin,
  Batanov, Becker, B\"{o}hringer, Bogomolov, Boller, Borm, Bornemann,
  Br\"{a}uninger, Br\"{u}ggen, Brunner, Brusa, Bulbul, Buntov, Burwitz,
  Burkert, Clerc, Churazov, Coutinho, Dauser, Dennerl, Doroshenko, Eder,
  Emberger, Eraerds, Finoguenov, Freyberg, Friedrich, Friedrich, F\"{u}rmetz,
  Georgakakis, Gilfanov, Granato, Grossberger, Gueguen, Gureev, Haberl,
  H\"{a}lker, Hartner, Hasinger, Huber, Ji, v.~Kienlin, Kink, Korotkov,
  Kreykenbohm, Lamer, Lomakin, Lapshov, Liu, Maitra, Meidinger, Menz, Merloni,
  Mernik, Mican, Mohr, M\"{u}ller, Nandra, Nazarov, Pacaud, Pavlinsky,
  Perinati, Pfeffermann, Pietschner, Ramos-Ceja, Rau, Reiffers, Reiprich,
  Robrade, Salvato, Sanders, Santangelo, Sasaki, Scheuerle, Schmid, Schmitt,
  Schwope, Shirshakov, Steinmetz, Stewart, Str\"{u}der, Sunyaev, Tenzer,
  Tiedemann, Tr\"{u}mper, Voron, Weber, Wilms, \& Yaroshenko}]{Predehl2021}
Predehl, P., Andritschke, R., Arefiev, V., {et~al.} 2021, Astronomy {\&}
  Astrophysics, 647, A1

\bibitem[{Prigozhin {et~al.}(2000)Prigozhin, Kissel, Bautz, Grant, LaMarr,
  Foster, \& Jr.}]{Prigozhin2000}
Prigozhin, G.~Y., Kissel, S.~E., Bautz, M.~W., {et~al.} 2000, in X-Ray and
  Gamma-Ray Instrumentation for Astronomy XI, ed. K.~A. Flanagan \& O.~H.~W.
  Siegmund, Vol. 4140, International Society for Optics and Photonics (SPIE),
  123 -- 134

\bibitem[{Puccio {et~al.}(2020)Puccio, Todaro, Cicero, Sciortino, Laurent,
  Ferrando, Giglia, Nannarone, \& Barbera}]{Puccio2020}
Puccio, E., Todaro, M., Cicero, U.~L., {et~al.} 2020, Journal of Astronomical
  Telescopes, Instruments, and Systems, 6

\bibitem[{{PV Lighthouse Pty. Ltd.}(Accessed on: 10/2021)}]{pvlighthouse}
{PV Lighthouse Pty. Ltd.} Accessed on: 10/2021, Refractive index library

\bibitem[{Reddy(1995)}]{Reddy1995}
Reddy, M.~R. 1995, Journal of Materials Science, 30, 281

\bibitem[{Rehr \& Albers(2000)}]{Rehr2000}
Rehr, J.~J. \& Albers, R.~C. 2000, Reviews of Modern Physics, 72, 621

\bibitem[{Rochus {et~al.}(2020)Rochus, Auch{\`{e}}re, Berghmans, Harra,
  Schmutz, Sch\"{u}hle, Addison, Appourchaux, Cuadrado, Baker, Barbay, Bates,
  BenMoussa, Bergmann, Beurthe, Borgo, Bonte, Bouzit, Bradley, B\"{u}chel,
  Buchlin, B\"{u}chner, Cab{\'{e}}, Cadiergues, Chaigneau, Chares, Cortez,
  Coker, Condamin, Coumar, Curdt, Cutler, Davies, Davison, Defise, Zanna,
  Delmotte, Delouille, Dolla, Dumesnil, D\"{u}rig, Enge, Fran{\c{c}}ois,
  Fourmond, Gillis, Giordanengo, Gissot, Green, Guerreiro, Guilbaud, Gyo,
  Haberreiter, Hafiz, Hailey, Halain, Hansotte, Hecquet, Heerlein, Hellin,
  Hemsley, Hermans, Hervier, Hochedez, Houbrechts, Ihsan, Jacques,
  J{\'{e}}r{\^{o}}me, Jones, Kahle, Kennedy, Klaproth, Kolleck, Koller,
  Kotsialos, Kraaikamp, Langer, Lawrenson, Clech', Lenaerts, Liebecq, Linder,
  Long, Mampaey, Markiewicz-Innes, Marquet, Marsch, Matthews, Mazy, Mazzoli,
  Meining, Meltchakov, Mercier, Meyer, Monecke, Monfort, Morinaud, Moron,
  Mountney, M\"{u}ller, Nicula, Parenti, Peter, Pfiffner, Philippon, Phillips,
  Plesseria, Pylyser, Rabecki, Ravet-Krill, Rebellato, Renotte, Rodriguez,
  Roose, Rosin, Rossi, Roth, Rouesnel, Roulliay, Rousseau, Ruane, Scanlan,
  Schlatter, Seaton, Silliman, Smit, Smith, Solanki, Spescha, Spencer, Stegen,
  Stockman, Szwec, Tamiatto, Tandy, Teriaca, Theobald, Tychon, van
  Driel-Gesztelyi, Verbeeck, Vial, Werner, West, Westwood, Wiegelmann, Willis,
  Winter, Zerr, Zhang, \& Zhukov}]{Rochus2020}
Rochus, P., Auch{\`{e}}re, F., Berghmans, D., {et~al.} 2020, Astronomy {\&}
  Astrophysics, 642, A8

\bibitem[{Rodgers {et~al.}(2015)Rodgers, Sorensen, Drolshagen, \&
  Santin}]{Rodgers2015}
Rodgers, D., Sorensen, J., Drolshagen, G., \& Santin, G. 2015, Athena
  environmental specification, {TEC-EES/2014.89/DR}, Tech. rep., ESA

\bibitem[{Salmaso {et~al.}(2021)Salmaso, Basso, Cotroneo, Ghigo, Pareschi,
  Redaelli, Sironi, Spiga, Tagliaferri, Vecchi, Fiorini, Incorvaia, Uslenghi,
  Paoletti, Ferrari, Zappettini, Lolli, del Rio, Burwitz, Christensen,
  Ferreira, Gellert, Massahi, Bavdaz, \& Ferreira}]{Salmaso2021}
Salmaso, B., Basso, S., Cotroneo, V., {et~al.} 2021, in Optics for {EUV},
  X-Ray, and Gamma-Ray Astronomy X, ed. G.~Pareschi, S.~L.
  O{\textquotesingle}Dell, \& J.~A. Gaskin ({SPIE})

\bibitem[{Saloman {et~al.}(1988)Saloman, Hubbell, \& Scofield}]{Saloman1988}
Saloman, E., Hubbell, J., \& Scofield, J. 1988, Atomic Data and Nuclear Data
  Tables, 38, 1

\bibitem[{Sciortino {et~al.}(2016)Sciortino, Cicero, Magnano, P{\'{\i}}{\v{s}},
  \& Barbera}]{Sciortino2016}
Sciortino, L., Cicero, U.~L., Magnano, E., P{\'{\i}}{\v{s}}, I., \& Barbera, M.
  2016, in Space Telescopes and Instrumentation 2016: Ultraviolet to Gamma Ray,
  ed. J.-W.~A. den Herder, T.~Takahashi, \& M.~Bautz ({SPIE})

\bibitem[{Serra {et~al.}(2017)Serra, Marvin, Moglie, Primiani, Cozza, Arnaut,
  Huang, Hatfield, Klingler, \& Leferink}]{Serra2017}
Serra, R., Marvin, A.~C., Moglie, F., {et~al.} 2017, {IEEE} Electromagnetic
  Compatibility Magazine, 6, 63

\bibitem[{Small \& Nix(1992)}]{Small1992}
Small, M.~K. \& Nix, W. 1992, Journal of Materials Research, 7, 1553

\bibitem[{{Smith M.J.S.}(2019)}]{Smith2019}
{Smith M.J.S.} 2019, Xmm-newton calibration technical note

\bibitem[{Sokolov {et~al.}(2014)Sokolov, Eggenstein, Erko, Follath,
  K\"{u}nstner, Mast, Schmidt, Senf, Siewert, Zeschke, \&
  Sch\"{a}fers}]{Sokolov2014}
Sokolov, A.~A., Eggenstein, F., Erko, A., {et~al.} 2014, in Advances in
  Metrology for X-Ray and {EUV} Optics V, ed. L.~Assoufid, H.~Ohashi, \& A.~K.
  Asundi ({SPIE})

\bibitem[{{Sopra S.A.}(Accessed on: 10/2021)}]{sopra}
{Sopra S.A.} Accessed on: 10/2021, Optical data from sopra sa

\bibitem[{Stern(1974)}]{Stern1974}
Stern, E.~A. 1974, Physical Review B, 10, 3027

\bibitem[{Stevie \& Donley(2020)}]{Stevie2020}
Stevie, F.~A. \& Donley, C.~L. 2020, Journal of Vacuum Science {\&} Technology
  A, 38, 063204

\bibitem[{Str\"{u}der {et~al.}(2001)Str\"{u}der, Briel, Dennerl, Hartmann,
  Kendziorra, Meidinger, Pfeffermann, Reppin, Aschenbach, Bornemann,
  Br\"{a}uninger, Burkert, Elender, Freyberg, Haberl, Hartner, Heuschmann,
  Hippmann, Kastelic, Kemmer, Kettenring, Kink, Krause, M\"{u}ller, Oppitz,
  Pietsch, Popp, Predehl, Read, Stephan, St\"{o}tter, Tr\"{u}mper, Holl,
  Kemmer, Soltau, St\"{o}tter, Weber, Weichert, von Zanthier, Carathanassis,
  Lutz, Richter, Solc, B\"{o}ttcher, Kuster, Staubert, Abbey, Holland, Turner,
  Balasini, Bignami, Palombara, Villa, Buttler, Gianini, Lain{\'{e}}, Lumb, \&
  Dhez}]{Struder2001}
Str\"{u}der, L., Briel, U., Dennerl, K., {et~al.} 2001, Astronomy {\&}
  Astrophysics, 365, L18

\bibitem[{Takahashi {et~al.}(2016)Takahashi, Kokubun, Mitsuda, Kelley, Ohashi,
  Aharonian, Akamatsu, Akimoto, Allen, Anabuki, Angelini, Arnaud, Asai, Audard,
  Awaki, Axelsson, Azzarello, Baluta, Bamba, Bando, Bautz, Bialas, Blandford,
  Boyce, Brenneman, Brown, Bulbul, Cackett, Canavan, Chernyakova, Chiao, Coppi,
  Costantini, de~Plaa, den Herder, DiPirro, Done, Dotani, Doty, Ebisawa,
  Eckart, Enoto, Ezoe, Fabian, Ferrigno, Foster, Fujimoto, Fukazawa, Furuzawa,
  Galeazzi, Gallo, Gandhi, Gilmore, Giustini, Goldwurm, Gu, Guainazzi, Haas,
  Haba, Hagino, Hamaguchi, Harayama, Harrus, Hatsukade, Hayashi, Hayashi,
  Hayashida, Hiraga, Hirose, Hornschemeier, Hoshino, Hughes, Ichinohe, Iizuka,
  Inoue, Inoue, Ishibashi, Ishida, Ishikawa, Ishimura, Ishisaki, Itoh, Iwata,
  Iyomoto, Jewell, Kaastra, Kallman, Kamae, Kara, Kataoka, Katsuda, Katsuta,
  Kawaharada, Kawai, Kawano, Kawasaki, Khangulyan, Kilbourne, Kimball, King,
  Kitaguchi, Kitamoto, Kitayama, Kohmura, Kosaka, Koujelev, Koyama, Koyama,
  Kretschmar, Krimm, Kubota, Kunieda, Laurent, Lebrun, Lee, Leutenegger,
  Limousin, Loewenstein, Long, Lumb, Madejski, Maeda, Maier, Makishima,
  Markevitch, Masters, Matsumoto, Matsushita, McCammon, McGuinness, McNamara,
  Mehdipour, Miko, Miller, Miller, Mineshige, Minesugi, Mitsuishi, Miyazawa,
  Mizuno, Mori, Mori, Moroso, Moseley, Muench, Mukai, Murakami, Murakami,
  Mushotzky, Nagano, Nagino, Nakagawa, Nakajima, Nakamori, Nakano, Nakashima,
  Nakazawa, Namba, Natsukari, Nishioka, Nobukawa, Nobukawa, Noda, Nomachi,
  O{\textquotesingle}Dell, Odaka, Ogawa, Ogawa, Ogi, Ohno, Ohta, Okajima,
  Okamoto, Okazaki, Ota, Ozaki, Paerels, Paltani, Parmar, Petre, Pinto, Pohl,
  Pontius, Porter, Pottschmidt, Ramsey, Reynolds, Russell, Safi-Harb, Saito,
  ichiro Sakai, Sakai, Sameshima, Sasaki, Sato, Sato, Sato, Sato, Sawada,
  Schartel, Serlemitsos, Seta, Shibano, Shida, Shidatsu, Shimada, Shinozaki,
  Shirron, Simionescu, Simmons, Smith, Sneiderman, Soong, Stawarz, Sugawara,
  Sugita, Sugita, Szymkowiak, Tajima, Takahashi, Takeda, Takei, Tamagawa,
  Tamura, Tamura, Tanaka, Tanaka, Tanaka, Tashiro, Tawara, Terada, Terashima,
  Tombesi, Tomida, Tsuboi, Tsujimoto, Tsunemi, Tsuru, Uchida, Uchiyama,
  Uchiyama, Ueda, Ueda, Ueno, Uno, Urry, Ursino, de~Vries, Wada, Watanabe,
  Watanabe, Werner, Wik, Wilkins, Williams, Yamada, Yamada, Yamaguchi, Yamaoka,
  Yamasaki, Yamauchi, Yamauchi, Yaqoob, Yatsu, Yonetoku, Yoshida, Yuasa,
  Zhuravleva, \& Zoghbi}]{Takahashi2016}
Takahashi, T., Kokubun, M., Mitsuda, K., {et~al.} 2016, in Space Telescopes and
  Instrumentation 2016: Ultraviolet to Gamma Ray, ed. J.-W.~A. den Herder,
  T.~Takahashi, \& M.~Bautz ({SPIE})

\bibitem[{{Tanaka} {et~al.}(2018){Tanaka}, {Uchida}, {Nakajima}, {Tsunemi},
  {Hayashida}, {Tsuru}, {Dotani}, {Nagino}, {Inoue}, {Katada}, {Washino},
  {Ozaki}, {Tomida}, {Natsukari}, {Ueda}, {Iwai}, {Mori}, {Yamauchi},
  {Hatsukade}, {Nishioka}, {Isoda}, {Nobukawa}, {Hiraga}, {Kohmura},
  {Murakami}, {Nobukawa}, {Bamba}, \& {Doty}}]{Tanaka2018}
{Tanaka}, T., {Uchida}, H., {Nakajima}, H., {et~al.} 2018, Journal of
  Astronomical Telescopes, Instruments, and Systems, 4, 011211

\bibitem[{{Tanaka} {et~al.}(1994){Tanaka}, {Inoue}, \& {Holt}}]{Tanaka1994}
{Tanaka}, Y., {Inoue}, H., \& {Holt}, S.~S. 1994, Publ. Astron. Soc. Jpn., 46,
  L37

\bibitem[{Tashiro {et~al.}(2020)Tashiro, Maejima, Toda, Kelley, Reichenthal,
  Hartz, Petre, Williams, Guainazzi, Costantini, Fujimoto, Hayashida,
  Henegar-Leon, Holland, Ishisaki, Kilbourne, Loewenstein, Matsushita, Mori,
  Okajima, Porter, Sneiderman, Takei, Terada, Tomida, Yamaguchi, Watanabe,
  Akamatsu, Arai, Audard, Awaki, Babyk, Bamba, Bando, Behar, Bialas,
  Boissay-Malaquin, Brenneman, Brown, Canavan, Chiao, Comber, Corrales, Cumbee,
  de~Vries, den Herder, Dercksen, Diaz-Trigo, DiPirro, Done, Dotani, Ebisawa,
  Eckart, Eckert, Eguchi, Enoto, Ezoe, Ferrigno, Fujita, Fukazawa, Furuzawa,
  Gallo, Gorter, Grim, Gu, Hagino, Hamaguchi, Hatsukade, Hawthorn, Hayashi,
  Hell, Hiraga, Hodges-Kluck, Horiuchi, Hornschemeier, Hoshino, Ichinohe, Iga,
  Iizuka, Ishida, Ishihama, Ishikawa, Ishimura, Jaffe, Kaastra, Kallman, Kara,
  Katsuda, Kenyon, Kimball, Kitaguti, Kitamoto, Kobayashi, Kobayashi, Kohmura,
  Kubota, Leutenegger, Li, Lockard, Maeda, Markevitch, Martz, Matsumoto,
  Matsuzaki, McCammon, McLaughlin, McNamara, Miko, Miller, Miller, Minesugi,
  Mitani, Mitsuishi, Mizumoto, Mizuno, Mukai, Murakami, Mushotzky, Nakajima,
  Nakamura, Nakazawa, Natsukari, Nigo, Nishioka, Nobukawa, Nobukawa, Noda,
  Odaka, Ogawa, Ohashi, Ohno, Ohta, Okamoto, Ota, Ozaki, Paltani, Plucinsky,
  Pottschmidt, Sampson, Sasaki, Sato, Sato, Sato, Sawada, Seta, Shibano, Shida,
  Shidatsu, Shigeto, Shinozaki, Shirron, Simionescu, Smith, Someya, Soong,
  Sugawara, Sugawara, Szymkowiak, Takahashi, Takeshima, Tamagawa, Tamura,
  Tanaka, Tanimoto, Terashima, Tsuboi, Tsujimoto, Tsunemi, Tsuru, Uchida,
  Uchida, Uchiyama, Ueda, Uno, Vink, Watanabe, Wittheof, Wolfs, Yamada,
  Yamaoka, Yamasaki, Yamauchi, Yamauchi, Yanagase, Yaqoob, Yasuda, Yoshida,
  Yoshioka, \& Zhuravleva}]{Tashiro2020}
Tashiro, M.~S., Maejima, H., Toda, K., {et~al.} 2020, in Space Telescopes and
  Instrumentation 2020: Ultraviolet to Gamma Ray, ed. J.-W.~A. den Herder,
  K.~Nakazawa, \& S.~Nikzad ({SPIE})

\bibitem[{Taylor {et~al.}(1981)Taylor, Andresen, Peacock, \& Zobl}]{Taylor1981}
Taylor, B.~G., Andresen, R.~D., Peacock, A., \& Zobl, R. 1981, in X-Ray
  Astronomy (Springer Netherlands), 479--494

\bibitem[{{Thayer} {et~al.}(2021){Thayer}, {Masterson}, {Allen}, {Ryu},
  {Bautz}, {Megerssa}, {Chodas}, {Guevel}, {Hoak}, {Hong}, {Lambert},
  {Grindlay}, \& {Binzel}}]{Thayer2021}
{Thayer}, C., {Masterson}, R., {Allen}, B., {et~al.} 2021, Journal of
  Astronomical Telescopes, Instruments, and Systems, 7, 046001

\bibitem[{Timmermans {et~al.}(2018)Timmermans, Mariano, Pollentier, Richard,
  Huyghebaert, \& Gallagher}]{Timmermans2018}
Timmermans, M.~Y., Mariano, M., Pollentier, I., {et~al.} 2018, Journal of
  Micro/Nanolithography, {MEMS}, and {MOEMS}, 17, 1

\bibitem[{Torma {et~al.}(2014)Torma, Kostamo, Sipila, Mattila, Kostamo,
  Kostamo, Lipsanen, Laubis, Scholze, Nelms, Shortt, \& Bavdaz}]{Torma2014}
Torma, P.~T., Kostamo, J., Sipila, H., {et~al.} 2014, {IEEE} Transactions on
  Nuclear Science, 61, 695

\bibitem[{Torma {et~al.}(2013)Torma, Sipila, Mattila, Kostamo, Kostamo,
  Kostamo, Lipsanen, Nelms, Shortt, Bavdaz, \& Laubis}]{Torma2013}
Torma, P.~T., Sipila, H.~J., Mattila, M., {et~al.} 2013, {IEEE} Transactions on
  Nuclear Science, 60, 1311

\bibitem[{Townsend {et~al.}(2016)Townsend, Senin, Blunt, Leach, \&
  Taylor}]{Townsend2016}
Townsend, A., Senin, N., Blunt, L., Leach, R., \& Taylor, J. 2016, Precision
  Engineering, 46, 34

\bibitem[{Trümper(1982)}]{Truemper1982}
Trümper, J. 1982, Advances in Space Research, 2, 241

\bibitem[{Tsuneta {et~al.}(1991)Tsuneta, Acton, Bruner, Lemen, Brown,
  Caravalho, Catura, Freeland, Jurcevich, Morrison, Ogawara, Hirayama, \&
  Owens}]{Tsuneta1991}
Tsuneta, S., Acton, L., Bruner, M., {et~al.} 1991, Solar Physics, 136, 37

\bibitem[{Turner {et~al.}(2001)Turner, Abbey, Arnaud, Balasini, Barbera,
  Belsole, Bennie, Bernard, Bignami, Boer, Briel, Butler, Cara, Chabaud, Cole,
  Collura, Conte, Cros, Denby, Dhez, Coco, Dowson, Ferrando, Ghizzardi,
  Gianotti, Goodall, Gretton, Griffiths, Hainaut, Hochedez, Holland, Jourdain,
  Kendziorra, Lagostina, Laine, Palombara, Lortholary, Lumb, Marty, Molendi,
  Pigot, Poindron, Pounds, Reeves, Reppin, Rothenflug, Salvetat, Sauvageot,
  Schmitt, Sembay, Short, Spragg, Stephen, Str\"{u}der, Tiengo, Trifoglio,
  Tr\"{u}mper, Vercellone, Vigroux, Villa, Ward, Whitehead, \&
  Zonca}]{Turner2001}
Turner, M. J.~L., Abbey, A., Arnaud, M., {et~al.} 2001, Astronomy {\&}
  Astrophysics, 365, L27

\bibitem[{Urayama {et~al.}(2012)Urayama, Fujii, Miyazaki, \&
  Kimoto}]{Urayama2012}
Urayama, F., Fujii, A., Miyazaki, E., \& Kimoto, Y. 2012, in Protection of
  Materials and Structures From the Space Environment (Springer Berlin
  Heidelberg), 235--242

\bibitem[{Vedder {et~al.}(1989)Vedder, Vallerga, Siegmund, Gibson, \&
  Hull}]{Vedder1989}
Vedder, P.~W., Vallerga, J.~V., Siegmund, O.~H., Gibson, J., \& Hull, J. 1989,
  in {EUV}, X-Ray, and Gamma-Ray Instrumentation for Astronomy and Atomic
  Physics, ed. C.~J. Hailey \& O.~H.~W. Siegmund ({SPIE})

\bibitem[{Villa {et~al.}(1998)Villa, Barbera, Collura, Palombara, Musso, Serio,
  Stillwell, Tognon, \& Turner}]{Villa1998}
Villa, G., Barbera, M., Collura, A., {et~al.} 1998, {IEEE} Transactions on
  Nuclear Science, 45, 921

\bibitem[{Vlassak \& Nix(1992)}]{Vlassak1992}
Vlassak, J. \& Nix, W. 1992, Journal of Materials Research, 7, 3242

\bibitem[{Wikus {et~al.}(2010)Wikus, Adams, Baker, Bandler, Brys, Dewey,
  Doriese, Eckart, Figueroa-Feliciano, Goeke, Hamersma, Hilton, Hwang, Irwin,
  Kelley, Kilbourne, Leman, McCammon, Okajima, R.~H.~O{\textquotesingle}Neal,
  Porter, Reintsema, Rutherford, Serlemitsos, Saab, Sato, Soong, \&
  Trowbridge}]{Wikus2010}
Wikus, P., Adams, J.~S., Baker, R., {et~al.} 2010, in Space Telescopes and
  Instrumentation 2010: Ultraviolet to Gamma Ray, ed. M.~Arnaud, S.~S. Murray,
  \& T.~Takahashi ({SPIE})

\bibitem[{Wu {et~al.}(2004)Wu, Chen, Du, Logan, Sippel, Nikolou, Kamaras,
  Reynolds, Tanner, Hebard, \& Rinzler}]{Wu2004}
Wu, Z., Chen, Z., Du, X., {et~al.} 2004, Science, 305, 1273

\bibitem[{Zabinsky {et~al.}(1995)Zabinsky, Rehr, Ankudinov, Albers, \&
  Eller}]{Zabinsky1995}
Zabinsky, S.~I., Rehr, J.~J., Ankudinov, A., Albers, R.~C., \& Eller, M.~J.
  1995, Physical Review B, 52, 2995

\bibitem[{Zhang {et~al.}(1998)Zhang, Lefever-Button, \& Powell}]{Zhang1998}
Zhang, Z.~M., Lefever-Button, G., \& Powell, F.~R. 1998, International Journal
  of Thermophysics, 19, 905

\bibitem[{Zombeck {et~al.}(1997)Zombeck, Barbera, Collura, \&
  Murray}]{Zombeck1997}
Zombeck, M.~V., Barbera, M., Collura, A., \& Murray, S.~S. 1997, The
  Astrophysical Journal, 487, L69

\bibitem[{Zombeck {et~al.}(1995)Zombeck, Chappell, Kenter, Moore, Murray,
  Fraser, \& Serio}]{Zombeck1995}
Zombeck, M.~V., Chappell, J.~H., Kenter, A.~T., {et~al.} 1995, in {EUV}, X-Ray,
  and Gamma-Ray Instrumentation for Astronomy {VI}, ed. O.~H.~W. Siegmund \&
  J.~V. Vallerga ({SPIE})

\bibitem[{Zombeck {et~al.}(1990)Zombeck, Conroy, F.~R.~Harnden, Roy,
  Braeuninger, Burkert, Hasinger, \& Predehl}]{Zombeck1990}
Zombeck, M.~V., Conroy, M.~A., F.~R.~Harnden, J., {et~al.} 1990, in {EUV},
  X-Ray, and Gamma-Ray Instrumentation for Astronomy, ed. O.~H.~W. Siegmund \&
  H.~S. Hudson ({SPIE})

\end{thebibliography}

\end{document}